\documentstyle[aps,preprint]{revtex}

\begin{document}
\title{On the self-consistent spin-wave theory of layered Heisenberg magnets}
\author{V.Yu. Irkhin, A.A. Katanin and M.I. Katsnelson}
\address{Institute of Metal Physics, 620219 Ekateriburg, Russia}
\maketitle

\begin{abstract}
The versions of the self-consistent spin-wave theories (SSWT) of
two-dimensional (2D) Heisenberg ferro- and antiferromagnets with a weak
interlayer coupling and/or magnetic anisotropy, that are based on the
non-linear Dyson-Maleev, Schwinger, and combined boson-pseudofermion
representations, are analyzed. Analytical results for the temperature
dependences of (sublattice) magnetization and short-range order parameter,
and the critical points are obtained. The influence of external magnetic
field is considered. Fluctuation corrections to SSWT are calculated within a
random-phase approximation which takes into account correctly leading and
next-leading logarithmic singularities. These corrections are demonstrated
to improve radically the agreement with experimental data on layered
perovskites and other systems. Thus an account of these fluctuations
provides a quantitative theory of layered magnets.
\end{abstract}

\pacs{75.10 Jm, 75.30.Gw, 75.70.Ak}

\section{Introduction}

Investigation of low-dimensional magnetism is an important branch of the
modern solid state physics. Experimental interest in this problem is
connected with the magnetic properties of copper-oxide high-$T_c$
superconductors, organic compounds, ferromagnetic films, multilayers and
surfaces \cite{allen}.

As stated by the Mermin-Wagner theorem, two-dimensional (2D) isotropic
magnets possess long-range order (LRO) only in the ground state. Unlike
purely 2D Heisenberg magnets, real layered compounds have finite values of
the magnetic ordering temperature $T_M\ll |J|$ ($J$ is the exchange
integral) due to weak interlayer coupling and/or magnetic anisotropy. The
smallness of transition temperature leads to some peculiar features of these
systems. When crossing $T_M,$ the short-range order (SRO) is not totally
destroyed (in the 2D situation it is maintained up to $T\sim |J|$), and a
broad region above $T_M$ with strong SRO exists. Corresponding experimental
indications are provided by the data on elastic and inelastic neutron
scattering: well-pronounced peaks of diffuse scattering were observed in La$%
_2$CuO$_4$ \cite{Birg1}, Rb$_2$MnF$_4$ and K$_2$NiF$_4$ \cite{Birg2}, and
well-defined spin waves in K$_2$MnF$_4$ up to $T\sim 2T_N$ \cite{Birg3}.

A great progress in the theory of the ground state and thermodynamic
properties was made with the use of rigorous mathematical methods (quantum
Monte-Carlo and renormalization group calculations). At the same time,
simple approaches, which yield an analytical description of a wide range of
physical properties, are very useful for practical purposes. At low
temperatures ($T\ll T_M$) the standard spin-wave theory works
satisfactorily. At higher temperatures corrections owing to spin-wave
interactions become important. These corrections were treated
self-consistently many years ago for 3D Heisenberg model in Ref. \cite{Loly}
The same results were obtained within a variational approach for isotropic 
\cite{BlochVar} and anisotropic \cite{Rast2} Heisenberg magnets.

For 2D magnets, close ideas were used recently by the ``boson mean-field
theory'' \cite{Arovas,Yosh} which is based on the representation of spin
operators through Schwinger bosons, and the ``modified spin-wave theory'' 
\cite{Tak} based on the Dyson-Maleev (DM) representation. Note that the
former approach differs drastically from the standard mean-field
approximation in the Heisenberg model: it takes into account spin-wave
excitations and is highly sensitive to the space dimensionality. The results
of these theories are in a good agreement with the scaling consideration 
\cite{Chakraverty} and experimental data on spin excitations in CuO$_2$
-planes. Being generalized to quasi-2D case (see, for example, \cite
{Sarker,Our1st,Liu}), these approaches lead to the same results as \cite
{Loly,BlochVar,Rast2}. The approaches of Refs. \cite{Arovas,Yosh,Tak} were
also applied to frustrated 2D \cite{Barab,OurFr,Xu,Oguch,Nish} and 3D \cite
{OurFr} antiferromagnets.

In the approaches \cite{Yosh,Tak} LRO is described in terms of boson
condensation, see also Ref.\cite{IK}. To continue the region of
applicability of the theory to disordered phase, a chemical potential of the
Bose system (fictious magnetic field) is introduced at $T>T_M,$ which is
determined from the condition of vanishing of (staggered) magnetization.
Introducing such a field can be more strictly justified within the
projection operator technique \cite{Chem}.

While the approach of Refs. \cite{Arovas,Yosh} corresponds to $N\rightarrow
\infty $ limit of the generalized $SU(N)$ Heisenberg model, the approach of
Ref. \cite{Tak} can be considered as the result of the self-consistent
first-order $1/S$-expansion, i.e. summation of all the bubble diagrams for
the self-energy (see also Ref. \cite{Loly}). As argued in the present paper,
this equivalence is preserved also between the first-order $1/N$ expansion
and second-order self-consistent $1/S$-expansion and seems to take place in
all the orders of perturbation theories discussed.

At the same time, above-discussed approaches (we refer them to as the
self-consistent spin-wave theories, SSWT) turn out to have some
shortcomings. The first one is mainly technical: the Bose condensation
picture is inapplicable for anisotropic systems, since they have a gap in
the excitation spectrum. Further, the $SU(N)$ symmetry in this case is
broken, so that the $1/N$-expansion in the $SU(N)$ model cannot be performed
in principle. However, as it was mentioned, this expansion (and also
description of LRO in terms of the Bose condensate) is not the only way to
SSWT.

The second shortcoming is much more essential. It concerns the description
of thermodynamics at temperatures that are comparable with $T_M$. In
particular, the description of the (sublattice) magnetization curve near the
ordering point is poor: the corresponding equation has two solutions, so
that $\overline{S}(T)$ does not vanish continuously (see, e.g., Refs., \cite
{Loly,Liu}). Besides that, the transition points are strongly overestimated.
These drawbacks are due to that at sufficiently high temperatures the
higher-order processes connected with dynamical interaction between spin
waves, and also kinematical interaction should be taken into account.

The kinematical interaction is important in a wide temperature region only
for systems where $T_M$ is not small in comparison with $|J|S^2$ (e.g., for
3D systems). This interaction can be, in principle, taken into account by
the projection operator technique (see, e.g., Ref. \cite{Chem}). However,
this technique is rather complicated and is not convenient for practical
purposes. Another way to obtain the corrections owing to the kinematical
interaction is the use of the Baryakhtar-Krivoruchko-Jablonsky (BKJ)
representation \cite{BKJr,BKJBook} of spin operators via bosons and
pseudofermions, which generalizes the DM representation (we do not know such
a generalization for the Schwinger bosons). The introduction of
pseudofermions gives in principle a possibility to exclude the contribution
of unphysical states. These pseudofermions can be easily incorporated into
the theory and, at least for 3D magnets, partially correct the
above-mentioned drawbacks of the early versions of SSWT.

For layered systems with $T_M\ll |J|S^2$ the kinematical interaction is less
important (in fact, it works only in a narrow critical region near $T_M$),
but higher-order (in the dynamic interaction) contributions should be
included. As will be shown in the present paper, an infinite RPA-type series
of diagrams are to be taken into account (as already mentioned, this is
equivalent to the first-order $1/N$-expansion in the $SU(N)$ model). Such a
procedure permits to describe the ``2D-like'' Heisenberg regime \cite
{Our1/N,OurRG} where thermodynamics is determined by fluctuations of 2D
Heisenberg nature. These results enable one to obtain the correct expression
for $T_M$ up to some non-singular constant in the denominator. At the same
time, the true critical region, where the spin-wave picture of the spectrum
is completely inadequate, turns out to be very narrow in the layered
systems. A satisfactory description of this region is possible within the $%
1/N$-expansion in the $O(N)$ model (see Refs. \cite{Our1/N,OurAnis}). The
latter model is based on a fluctuation rather than spin-wave picture of
excitation spectrum. This circumstance provides important advantages at high
temperatures, but leads to some difficulties at the description of the low-
and intermediate-temperature regions. Therefore the approach based on the $%
SU(N)$ model ($1/S$-expansion) is more appropriate at not too high
temperatures.

In the present paper we formulate a version of the SSWT, which is to a large
measure free from above-mentioned shortcomings and is a good starting point
for further improvements. To this end we (i) use the BKJ representation to
obtain the correct description at not too low temperatures (ii) discard the
condition $\overline{S}=0$ and do not describe LRO in terms of Bose
condensate, which permits to treat anisotropic systems. We also demonstrate
(where possible) how our results can be obtained by the Schwinger boson
method \cite{Yosh}. Further we calculate the corrections to SSWT for
quasi-2D and easy-axis 2D magnets with small interlayer coupling or
anisotropy using second-order spin-wave results in the self-consistent form.

The plan of the present paper is as follows. In Sect.\ref{Reprs} we describe
the representation of spin operators by Schwinger and
Baryakhtar-Krivoruchko-Yablonsky \cite{BKJr}. In Sects. \ref{Quasi-2D} and 
\ref{E-ax} we consider thermodynamics of quasi-2D and anisotropic 2D layered
magnets within SSWT and construct an interpolation scheme between 2D and 3D
cases. In Sect. \ref{Field} we treat the problem of introducing magnetic
field into SSWT and calculating magnetic susceptibility. In Sect. \ref
{Interact} we investigate fluctuation corrections to the SSWT results, in
particular to the (sublattice) magnetization and ordering temperature, and
compare the results of our calculations with experimental data.

\section{Boson representations of the spin operators}

\label{Reprs}We consider the anisotropic Heisenberg model 
\begin{equation}
H=-\frac 12\sum_{ij}J_{ij}{\bf S}_i{\bf S}_j-\frac 12\eta
\sum_{ij}J_{ij}S_i^zS_j^z-D\sum_i(S_i^z)^2  \label{H}
\end{equation}
where $J_{ij}$ are the exchange integrals, $\eta >0$ and $D>0$ are the
two-site and single-site easy-axis anisotropy parameters.

Consider first the Baryakhtar-Krivoruchko-Jablonsky representation \cite
{BKJr,BKJBook} 
\begin{eqnarray}
S_i^{+} &=&\sqrt{2S}b_i\,,\;S_i^z=S-b_i^{\dagger }b_i-(2S+1)c_i^{\dagger }c_i
\label{BKJ} \\
S_i^{-} &=&\sqrt{2S}(b_i^{\dagger }-\frac 1{2S}b_i^{\dagger }b_i^{\dagger
}b_i)-\frac{2(2S+1)}{\sqrt{2S}}b_i^{\dagger }c_i^{\dagger }c_i  \nonumber
\end{eqnarray}
where $b_i^{\dagger },b_i\ $are the Bose ideal magnon operators, and $%
c_i^{\dagger },c_i\ $are the auxiliary pseudofermion operators at the site $%
i $ which take into account the kinematic interaction of spin waves. For the
states $|p\rangle $ in the physical subspace (with the boson occupation
numbers $N_i<2S$ and pseudofermion occupation numbers $F_i=0$) we have $%
c|p\rangle =0,$ and the representation (\ref{BKJ}) reduces to the standard
DM representation. The states $|u_0\rangle $ with $N_i>2S,$ $F_i=0$ and $%
|u_1\rangle $ with $F_i=1$ are unphysical. As shown in Ref. \cite{BKJr}, the
partition function can be calculated as 
\begin{equation}
{\cal Z}=\text{Sp}\left\{ \exp \left( -\beta H[b,b^{\dagger },c,c^{\dagger
}]-i\pi \sum_ic_i^{\dagger }c_i\right) \right\}  \label{zp}
\end{equation}
where $H[b,b^{\dagger },c,c^{\dagger }]$ is the original spin Hamiltonian (%
\ref{H}) written through the Bose and Fermi operators according to (\ref{BKJ}%
). Analogous relations take place for the averages of spin operators. It can
be proved\cite{BKJr} that the contribution of states $|u_0\rangle $ in (\ref
{zp}) is exactly canceled by the contribution of the states $|u_1\rangle $

Thus, unlike the DM representation, using the BKJ representation gives a
possibility to exclude the contribution of the states with the boson
occupation numbers $N_i>2S$ to thermodynamic quantities. It should be noted
that this property relates to the exact Hamiltonian of boson-pseudofermion
system $H[b,b^{\dagger },c,c^{\dagger }]$ only and does not necessarily hold
for its approximate expressions. However, one could expect that the
introduction of the Fermi operators extends the region of applicability of
approximate methods to not too low temperatures.

The factor $\exp \left( -i\pi \sum_ic_i^{\dagger }c_i\right) $ in (\ref{zp})
results in that the distribution function of the pseudofermions becomes $%
-N(E_f)$ where $N(E)=1/[\exp (E/T)-1]$ is the Bose function, $E_f$ is the
excitation energy for pseudofermions (as follows from the representation (%
\ref{BKJ}), the $c$-field has no dispersion).

In the case of a two-sublattice antiferromagnet we separate the lattice into 
$A$ and $B$ sublattices. On the sublattice $A$ we use the representation
that is similar to (\ref{BKJ}) 
\begin{eqnarray}
S_i^{+} &=&\sqrt{2S}a_i\,,\;S_i^z=S-a_i^{\dagger }a_i-(2S+1)c_i^{\dagger
}c_i,\;i\in A  \label{BKJa} \\
S_i^{-} &=&\sqrt{2S}(a_i^{\dagger }-\frac 1{2S}a_i^{\dagger }a_i^{\dagger
}a_i)-\frac{2(2S+1)}{\sqrt{2S}}a_i^{\dagger }c_i^{\dagger }c_i  \nonumber
\end{eqnarray}
and on the sublattice $B$ the ``conjugate'' representation: 
\begin{eqnarray}
S_i^{+} &=&\sqrt{2S}b_i^{\dagger }\,,\;S_i^z=-S+b_i^{\dagger
}b_i^{}+(2S+1)d_i^{\dagger }d_i^{},\;i\in B  \label{BKJb} \\
S_i^{-} &=&\sqrt{2S}(b_i^{}-\frac 1{2S}b_i^{\dagger }b_i^{}b_i^{})-\frac{%
2(2S+1)}{\sqrt{2S}}d_i^{\dagger }d_i^{}b_i^{}  \nonumber
\end{eqnarray}
where $a_i^{\dagger },a_i,$ and $b_i^{\dagger },b_i$ are the Bose operators, 
$c_i^{\dagger },c_i,$ and $d_i^{\dagger },d_i$ are the Fermi operators.

Another useful representation of spin operators is the Schwinger-boson
representation 
\begin{equation}
{\bf S}_i=\sum_{\sigma \sigma ^{\prime }}s_{i\sigma }^{\dagger }{\bbox %
\sigma }_{\sigma \sigma ^{\prime }}s_{i\sigma ^{\prime }}
\end{equation}
where ${\bbox \sigma }$ are the Pauli matrices, $\sigma ,\sigma ^{\prime
}=\uparrow ,\downarrow ,$ so that 
\begin{equation}
S_i^z=\frac 12(s_{i\uparrow }^{\dagger }s_{i\uparrow }-s_{i\downarrow
}^{\dagger }s_{i\downarrow }),\,\,S_i^{+}=s_{i\uparrow }^{\dagger
}s_{i\downarrow },\,\,S_i^{-}=s_{i\downarrow }^{\dagger }s_{i\uparrow }
\label{sb}
\end{equation}
The constraint condition 
\begin{equation}
s_{i\uparrow }^{\dagger }s_{i\uparrow }{}+s_{i\downarrow }^{\dagger
}s_{i\downarrow }=2S  \label{ccond}
\end{equation}
should be satisfied at each lattice site. Since the phases of $s_{i\uparrow
}\,$ and $s_{i\downarrow }$ can be simultaneously changed, $s_{i\sigma
}\rightarrow s_{i\sigma }\exp (i\phi _i),$ this representation possesses a
gauge symmetry. The Schwinger-boson representation can be simply related
with the Holstein-Primakoff representation if we fix the gauge by the
condition of hermiticity for one of the operators $s_{i\sigma }$, say, $%
s_{i\uparrow }$, i.e. $s_{i\uparrow }^{\dagger }=s_{i\uparrow }$. Then we
have from (\ref{ccond}) 
\begin{equation}
s_{i\uparrow }=\sqrt{2S-s_{i\downarrow }^{\dagger }s_{i\downarrow }},
\end{equation}
and substituting this into (\ref{sb}) we obtain the Holstein-Primakoff
representation. Thus the representations of the Schwinger bosons and by
Holstein-Primakoff are equivalent. As well as for the BKJ representation,
this equivalence can be violated in approximate treatments. Unlike the
Holstein-Primakoff (or DM) representation, the Schwinger-boson
representation can be easily generalized to an arbitrary number of boson
``flavors'' $N\geq 2,$ and $1/N$-expansion can be developed. At the same
time, there is no natural way to take into account ``roughly'' the
kinematical interaction by introducing the Fermi operators into this
representation .

In the antiferromagnetic case we pass following to Ref. \cite{Arovas} to the
local coordinate system by the replacement 
\[
s_{i\uparrow }\rightarrow -s_{i\downarrow },\,\,s_{i\downarrow }\rightarrow
s_{i\uparrow } 
\]
at one of two sublattices.

\section{SSWT of quasi-2D magnets}

\label{Quasi-2D}

\subsection{Self-consistent approach within the BKJ representation}

\label{Q2dBKJ}In this section we consider the quasi-2D case with $D=\eta
_{ij}=0$ and the exchange integrals $J_{ij}=J$ for $i,j$ being nearest
neighbors in the same plane and $J_{ij}=J^{\prime }$ for $i,j$ in different
planes. First we use the BKJ representation. In the ferromagnetic case the
Heisenberg Hamiltonian (\ref{H})\ takes the form 
\begin{eqnarray}
H &=&-\frac 12\sum_{ij}J_{ij}\left[ (S-b_i^{\dagger }b_i-(2S+1)c_i^{\dagger
}c_i)(S-b_j^{\dagger }b_j-(2S+1)c_j^{\dagger }c_j)\right.  \nonumber \\
&&\ \ \ \ \ \left. +2S(b_i^{\dagger }-\frac 1{2S}b_i^{\dagger }b_i^{\dagger
}b_i)b_j-2(2S+1)b_i^{\dagger }b_jc_i^{\dagger }c_i\right]  \label{HFQ} \\
&&\ \ \ \ \ -\mu \sum_i\left[ b_i^{\dagger }b_i+(2S+1)c_i^{\dagger
}c_i\right]  \nonumber
\end{eqnarray}
To satisfy the condition $\overline{S}=0$ in the paramagnetic phase we have
introduced the Lagrange multiplier $\mu $. This multiplier corresponds to
the constraint of the total number of bosons and pseudofermions at $T>T_C$
and plays a role of common chemical potential $\mu $ of the
boson-pseudofermion system (for the pure boson system it was introduced in
Refs. \cite{Rastelly1,Tak}). At $T<T_C$ we have $\mu =0$ since no
restriction of boson and pseudofermion occupation numbers is needed here.
Introducing the chemical potential, which gives a possibility to continue
the theory into the disordered phase, can be justified more strictly if one
takes into account the kinematical interaction in a regular way\cite{Chem}.
Since the magnon number is not conserved at $T<T_C,$ the Bose condensation
which takes place in Refs. \cite{Yosh,Tak} does not occur in our approach.

Further we perform decouplings of the quartic forms which occur in (\ref{HFQ}%
). Introducing the averages 
\begin{equation}
\gamma =\overline{S}+\langle b_i^{\dagger }b_{i+\delta _{\perp }}\rangle
,\,\,\,\,\gamma ^{\prime }=\overline{S}+\langle b_i^{\dagger }b_{i+\delta
_{\parallel }}\rangle  \label{ksi}
\end{equation}
we derive the quadratic Hamiltonian of the mean-field approximation in the
form 
\begin{eqnarray}
H &=&\sum_{i\delta }J_\delta \gamma _\delta \,\left[ b_i^{\dagger
}b_i-b_{i+\delta }^{\dagger }b_i+(2S+1)c_i^{\dagger }c_i\right]  \nonumber \\
&&\ \ \ \ \ \ \ \ -\mu \sum_i\left[ b_i^{\dagger }b_i+(2S+1)c_i^{\dagger
}c_i\right]
\end{eqnarray}
where $\gamma _{\delta _{\perp }}=\gamma $ and $\gamma _{\delta _{\parallel
}}=\gamma ^{\prime }.$ From the definition of $\gamma $ (\ref{ksi}) one
finds the system of self-consistent equations 
\begin{equation}
\gamma =\overline{S}+\sum_{{\bf k}}N_{{\bf k}}\cos k_x,\,\,\gamma ^{\prime }=%
\overline{S}+\sum_{{\bf k}}N_{{\bf k}}\cos k_z,  \label{gg1}
\end{equation}
which should be solved together with the condition 
\begin{equation}
\overline{S}=S+(2S+1)N(E_f)-\sum_{{\bf k}}N_{{\bf k}}  \label{MagnF}
\end{equation}
where $N_{{\bf k}}=N(E_{{\bf k}})$ are the Bose occupation numbers, 
\begin{eqnarray}
E_f &=&(2S+1)(\Gamma _0-\mu ) \\
E_{{\bf k}} &=&\Gamma _0-\Gamma _{{\bf k}}-\mu  \nonumber
\end{eqnarray}
are the pseudofermion excitation energy and the spin-wave spectrum
respectively, 
\begin{equation}
\Gamma _{{\bf k}}=2\left[ |J|\gamma (\cos k_x+\cos k_y)+|J^{\prime }|\gamma
^{\prime }\cos k_z\right]
\end{equation}

Consider now the case of an antiferromagnet. Introducing the operators 
\begin{equation}
B_i=\left\{ 
\begin{array}{cc}
a_i & i\in A \\ 
b_i^{\dagger } & i\in B
\end{array}
\right. ,\,\,\,\,\,C_i=\left\{ 
\begin{array}{cc}
c_i & i\in A \\ 
d_i & i\in B
\end{array}
\right.  \label{BC}
\end{equation}
we derive 
\begin{eqnarray}
H_{AF} &=&|J|\gamma \sum_{i,\delta _{\perp }}\left[ B_i^{\dagger
}B_i^{}-B_{i+\delta }^{\dagger }B_i^{}+(2S+1)C_i^{\dagger }C_i^{}\right]
\label{HAF} \\
&&\ \ \ \ \ \ \ \ \ \ \ \ \ +|J^{\prime }|\gamma ^{\prime }\sum_{i,\delta
_{\parallel }}\left[ B_i^{\dagger }B_i^{}-B_{i+\delta }^{\dagger
}B_i+(2S+1)C_i^{\dagger }C_i^{}\right]  \nonumber \\
&&\ \ \ \ \ \ \ \ \ \ \ \ \ -\mu \sum_i\left[ B_i^{\dagger
}B_i^{}+(2S+1)C_i^{\dagger }C_i\right]  \nonumber
\end{eqnarray}
where 
\begin{equation}
\gamma =\overline{S}+\langle a_ib_{i+\delta _{\perp }}\rangle ,\,\gamma
^{\prime }=\overline{S}+\langle a_ib_{i+\delta _{\parallel }}\rangle
\label{ksia}
\end{equation}
Diagonalizing this Hamiltonian one finds the self-consistent equations 
\begin{eqnarray}
\gamma &=&\overline{S}+\sum_{{\bf k}}\frac{\Gamma _{{\bf k}}}{2E_{{\bf k}}}%
\cos k_x\coth \frac{E_{{\bf k}}}{2T}  \label{pureAF} \\
\gamma ^{\prime } &=&\overline{S}+\sum_{{\bf k}}\frac{\Gamma _{{\bf k}}}{2E_{%
{\bf k}}}\cos k_z\coth \frac{E_{{\bf k}}}{2T}  \nonumber \\
\overline{S} &=&(S+1/2)\coth \frac{E_f}{2T}-\sum_{{\bf k}}\frac{\Gamma
_0-\mu }{2E_{{\bf k}}}\coth \frac{E_{{\bf k}}}{2T}  \nonumber
\end{eqnarray}
where the antiferromagnetic spin-wave spectrum has the form 
\begin{equation}
E_{{\bf k}}=\sqrt{(\Gamma _0-\mu )^2-\Gamma _{{\bf k}}^2}
\end{equation}
with $\Gamma _{{\bf k}}$ and $E_f$ being the same as in the ferromagnetic
case.

For both ferro- and antiferromagnetic cases, the calculation of spin
correlation functions \cite{Tak} shows that $\mu $ is directly connected
with the correlation length $\xi _\delta $ in the direction $\delta $ by the
relation 
\begin{equation}
\xi _\delta ^{-1}=\sqrt{-\mu /|J_\delta \gamma _\delta |}  \label{corr}
\end{equation}
The parameters $\gamma $ and $\gamma ^{\prime }$ are also simply related to
the spin correlation function at the nearest-neighbor sites, 
\begin{equation}
|\langle {\bf S}_i{\bf S}_{i+\delta }\rangle |=\gamma _\delta ^2,
\label{CFF}
\end{equation}
and therefore play a role of SRO parameters. For the total energy we readily
obtain 
\begin{equation}
{\cal E}=-\frac 12\sum_\delta |J_{i,i+\delta }|\gamma _\delta
^2=-(2|J|\gamma ^2+|J^{\prime }|\gamma ^{\prime 2})  \label{EF}
\end{equation}
In the classical limit $S\rightarrow \infty $ the SSWT equations are
simplified. Supposing $T\gg |J|S$ ($T_M\sim |J|S^2$ in this case) the
equations for both FM and AFM cases reduce to 
\begin{eqnarray}
\overline{S}/S &=&\coth (E_f/2T)-\frac TS\sum_{{\bf k}}\frac 1{\Gamma
_0-\Gamma _{{\bf k}}-\mu }  \nonumber \\
\gamma  &=&\overline{S}+T\sum_{{\bf k}}\frac{\cos k_x}{\Gamma _0-\Gamma _{%
{\bf k}}-\mu },\,  \nonumber \\
\,\gamma ^{\prime } &=&\overline{S}+T\sum_{{\bf k}}\frac{\cos k_z}{\Gamma
_0-\Gamma _{{\bf k}}-\mu },
\end{eqnarray}
For $T<T_M$ ($\mu =0$) the averaged (over nearest neighbors) SRO parameter 
\begin{equation}
\gamma _{\text{ef}}(T)=(4J\gamma +2J^{\prime }\gamma ^{\prime })/J_0
\label{BG}
\end{equation}
(but not the magnetization) satisfies the standard mean-field equation 
\begin{equation}
\gamma _{\text{ef}}/S=B_\infty \left( J_0\gamma _{\text{ef}}S/T\right) 
\label{BB}
\end{equation}
where $B_\infty (x)=\coth x-1/x$ is the classical Brillouin function
(Langevin function). The temperature $T^{*}$ where $\gamma _{\text{ef}%
}(T^{*})=0$ is higher than $T_M$, so that we have $\gamma _{ef}(T_M)>0$ and
the behavior of $\gamma _{ef}$ for $T>T_M$ is more complicated in comparison
with (\ref{BB}).

\subsection{Approximation of effective SRO parameter}

\label{Interp}Equations (\ref{gg1}), (\ref{MagnF}) and (\ref{pureAF}) still
demonstrate unphysical behavior of magnetization for $T$ close to $T_M$ at
small enough $J^{\prime }/J\ $(see below). Introducing the pseudofermion
field does not improve situation in this case: the transition temperature is
already too small to be influenced by pseudofermion excitations with the
energy of order of $|J|$. As discussed in the Introduction, the dynamical
spin-wave interaction should be treated more correctly in such a situation.
A rough solution of this problem can be achieved by the replacement 
\begin{equation}
\sum_\delta J_{i,i+\delta }\gamma _\delta (b_i^{\dagger }b_i-b_i^{\dagger
}b_{i+\delta })\rightarrow \gamma _{\text{ef}}\sum_\delta J_{i,i+\delta
}(b_i^{\dagger }b_i-b_i^{\dagger }b_{i+\delta })  \label{Approx}
\end{equation}
where $\gamma _{\text{ef}}$ is determined by (\ref{BG}). Then we obtain the
spectrum 
\begin{eqnarray}
E_{{\bf q}} &=&\gamma _{\text{ef}}(J_0-J_{{\bf q}})-\mu ,\;\text{FM} 
\nonumber \\
E_{{\bf q}} &=&\sqrt{(J_0\gamma _{\text{ef}}-\mu )^2-(J_{{\bf q}}\gamma _{%
\text{ef}})^2},\;\text{AFM}  \label{EqI}
\end{eqnarray}
and the pseudofermion excitation energy 
\begin{equation}
E_f=(2S+1)(\gamma _{\text{ef}}J_0-\mu ),
\end{equation}
(here and hereafter we use the definition $J_{{\bf q}}=\sum_\delta |J_\delta
|\exp (i{\bf q}{\bbox \delta )}$). The SSWT equations take the form 
\begin{eqnarray}
\overline{S} &=&S+(2S+1)N(E_f)-\sum_{{\bf k}}N_{{\bf k}}  \label{EqIF} \\
\gamma _{\text{ef}} &=&\overline{S}+\frac 1{J_0}\sum_{{\bf k}}J_{{\bf k}}N_{%
{\bf k}}  \nonumber
\end{eqnarray}
in the FM case and 
\begin{eqnarray}
\overline{S} &=&(S+1/2)\coth \frac{E_f}{2T}-\gamma _{\text{ef}}\sum_{{\bf k}}%
\frac{J_0}{2E_{{\bf k}}}\coth \frac{E_{{\bf k}}}{2T}  \label{EqIAF} \\
\gamma _{\text{ef}} &=&\overline{S}+\frac{\gamma _{\text{ef}}}{J_0}\sum_{%
{\bf k}}\frac{J_{{\bf k}}^2}{2E_{{\bf k}}}\coth \frac{E_{{\bf k}}}{2T} 
\nonumber
\end{eqnarray}
in the AFM case. The approximation (\ref{Approx}) is analogous to passing
from the Hartree-Fock approximation to the local approximation in the
spin-density functional method for the electron gas. As it is known, due to
account of screening effects, such approximations can lead to improving the
results and eliminating the unphysical peculiarities. Note that the same
equations (\ref{EqIF}) and (\ref{EqIAF})\ were obtained earlier in Ref. \cite
{Antz}. However, when deriving these equations, the authors have used
expressions for the spin Green's function which have incorrect $\omega
\rightarrow \infty $ asymptotics.

Another approach, which gives a possibility to improve the behavior of
(sublattice) magnetization near $T_M,$ is based on a variational principle
and is considered in Appendix A. It leads to the same spin-wave spectrum (%
\ref{EqI}), but the corresponding SSWT equations are somewhat different from
(\ref{EqIF}) and (\ref{EqIAF}). However, numerical calculations shows that
this difference is very small (several percents of magnetization value), and
further we will refer to both these approaches as the approximation of
effective SRO parameter.

\subsection{Temperature dependences of long- and short-range order parameters
}

\label{TempDep}In the two-dimensional case $J^{\prime }=0$ the spectrum $E_{%
{\bf k}}$ is independent of $\gamma ^{\prime },$ and two remaining equations
for $\overline{S}$ and $\gamma $ differ from those of approaches of Refs. 
\cite{Arovas,Sarker} only by the presence of pseudofermion distribution
function $N(E_f)$ which describes the kinematical interaction of spin waves.
At $T=0$ we have $\mu =0,$ and $\gamma =\overline{S}=S$ in FM case and $%
\gamma >S,$ $\overline{S}<S$ in AFM case, which corresponds to magnetic
ordering in the ground state. As well as in Refs. \cite{Arovas,Sarker},
equations (\ref{gg1}), (\ref{MagnF}) and (\ref{pureAF}) do not have at $T>0$
solutions with $\mu =0,\;\overline{S}>0$ since the integrals in equations (%
\ref{gg1}), (\ref{MagnF}) and (\ref{pureAF}) are logarithmically divergent
in this case, and the only solution of these equations for $J^{\prime }=0$
is $\overline{S}=0,$ $\mu <0,$ which corresponds to a disordered phase.

At low temperatures $T\ll |J|S^2$ we can neglect the pseudofermion
contribution (i.e. kinematical interaction of spin waves) and we completely
reproduce the results of Refs.\cite{Arovas,Yosh,Tak,Sarker}. In particular,
the correlation length has the exponential dependence 
\begin{mathletters}
\begin{eqnarray}
\xi &=&C_\xi ^{\text{F}}\exp \left( 2\pi JS^2/T\right) \text{\ \ \ (FM)}
\label{CLFM} \\
\xi &=&C_\xi ^{\text{AF}}\exp \left( 2\pi |J|\gamma _0\overline{S}%
_0/T\right) \text{ (AFM)}  \label{CLAFM}
\end{eqnarray}
where 
\end{mathletters}
\begin{equation}
\overline{S}_0=S-0.1971,\text{ }\gamma _0=S+0.079  \label{D2}
\end{equation}
are the 2D ground-state LRO and SRO parameters, $C_\xi ^{\text{F,AF}}$ are
the constants. The result (\ref{CLAFM}) was obtained earlier within the
one-loop RG approach \cite{Chakraverty}. With increasing $T$ the role of
kinematical interaction increases and for $T\sim JS^2$ we cannot neglect $%
N(E_f)$. The dependence $\gamma (T)$ for $J^{\prime }=0$ is shown and
compared with the result of approaches \cite{Yosh,Tak} in Fig.\ref{FigSROq2D}
Unlike the approaches \cite{Yosh,Tak}, equations (\ref{gg1}), (\ref{MagnF})
and (\ref{pureAF}) do not lead to the non-physical phase transition with
vanishing of the SRO parameter, and the latter is finite at any
temperatures. Note that for $J^{\prime }=0$ the equations (\ref{EqI}) and (%
\ref{EqIAF}) give the same results.

In the presence of interlayer coupling, the integrals in the SSWT equations (%
\ref{gg1}), (\ref{MagnF}) and (\ref{pureAF}) becomes convergent at finite $T$
even at $\mu =0.$ For not too high temperatures $T<T_M$ (the ordering
temperature $T_M$ will be calculated below) these equations have the
solution with $\overline{S}>0,$ which corresponds to the ordered magnetic
phase$.$ For $T>T_M$ we again have $\overline{S}=0$ and $\mu <0$ as well as
in 2D case at finite $T.$

Figs.\ref{FigSROq2D}-\ref{FigMu} show the results of the numerical solution
of the equations of Sects. \ref{Q2dBKJ}, \ref{Interp} for different values
of the interlayer coupling. In the three-dimensional case ($J^{\prime }\ =J$%
) the (staggered) magnetization vanishes at $T_C=1.20J$ ($T_N=1.33|J|$) that
is approximately by 20\% higher than the corresponding value obtained from
the high-temperature series expansion. At the same time, the ratio $%
T_N/T_C=1.20$ is in agreement with the results of this expansion. The SRO
parameter $\gamma $ demonstrates a sharp decrease in a narrow temperature
region above $T_M$ and then asymptotically goes to zero. One can see that $%
\gamma _{\text{AFM}}>\gamma _{\text{FM}}$ due to the quantum fluctuations.
At the transition point we have $\gamma _c\equiv \gamma (T_M)=0.62$ for FM
case and $\gamma _c=0.70$ for AFM case. The dependences $\gamma (T)/S$ for
3D ferromagnets with different $S$ are shown in Fig. \ref{FigSROs} One can
see that the value $\gamma _c/S$ rapidly decreases with increasing $S$,
reaching $\gamma _c=0.39$ at $S\rightarrow \infty .$ Thus at $S=1/2$ strong
quantum fluctuations are present even at $T=T_M$.

Consider now the quasi-2D case $0<J^{\prime }/J<1.$ At $J^{\prime }/J<0.4$
equations (\ref{gg1}), (\ref{MagnF}) and (\ref{pureAF}) still yield
unphysical behavior of magnetization and SRO parameters for $T$ close to $%
T_M $ (as shown on Figs. \ref{FigSROq2D} and \ref{FigMagn} for $J^{\prime
}/J=0.3 $). At the same time, the approximation of single effective SRO
parameter improves the behavior of magnetization for $T$ close to $T_M$ and
provides a qualitatively correct description of thermodynamics at arbitrary
temperatures$.$ The price which we pay is overestimation of $T_M$ even in
comparison with the results of Eqs. (\ref{gg1}), (\ref{MagnF}) and (\ref
{pureAF}), since the temperature dependence of the ratio of effective inter-
and intralayer couplings (which is $J^{\prime }/J$ for spectrum (\ref{EqI}))
is absent in the approximation used. In particular, for $J^{\prime
}/J\rightarrow 0$ the results obtained within approximation (\ref{Approx})
are different from those of standard spin-wave theory only by quantum
(ground-state) renormalization of $\gamma $. Note that according to Fig.\ref
{FigSROq2D} with decreasing $J^{\prime }/J$ the size of the region with
noticeable SRO increases.

At small $T-T_M$ we have $-\mu \propto (T-T_M)^2$ (see Fig. \ref{FigMu} for
a ferromagnetic case, the same situation takes place in the AFM case) so
that, according to (\ref{corr}), the critical exponent for the correlation
length is $\nu =1$. Since the magnetization changes linearly near $T_M,$ we
have also $\beta =1.$ The influence of higher-order terms in $1/S$ on these
results is discussed in Sect. \ref{Interact}. Note that if we determine,
following to Ref.\cite{Nag}, the critical exponent $\nu $ from a not too
narrow temperature interval near $T_M$ $,$ this becomes more close to the
experimental value.

At very low temperatures ($T\ll |J^{\prime }|S$) and arbitrary $J^{\prime
}/J $ the calculation can be performed analytically. The corrections to
magnetization of a ferromagnet are proportional to $T^{3/2}$%
\begin{equation}
\overline{S}=S-\frac 1{8\pi ^{3/2}}\sqrt{\frac J{J^{\prime }}}\left( \frac T{%
JS}\right) ^{3/2}\zeta (3/2)
\end{equation}
where $\zeta (3/2)$ is the Riemann zeta-function. At the same time, SRO
parameters have a more weak $T^{5/2}$-dependence 
\begin{eqnarray}
\gamma &=&S-\frac 3{32\pi ^{3/2}}\sqrt{\frac J{J^{\prime }}}\left( \frac T{JS%
}\right) ^{5/2}\zeta (5/2) \\
\gamma ^{\prime } &=&S-\frac 3{32\pi ^{3/2}}\left( \frac J{J^{\prime }}%
\right) ^{3/2}\left( \frac T{JS}\right) ^{5/2}\zeta (5/2)
\end{eqnarray}
For $J^{\prime }=J$ this result corresponds to that of the Dyson theory \cite
{Dyson} to leading order in $1/S$. For an antiferromagnet we have 
\begin{equation}
\overline{S}=\overline{S}_0-\frac{T^2}{24c\sqrt{JJ^{\prime }\gamma _0\gamma
_0^{\prime }}}
\end{equation}
where $\gamma _0,\gamma _0^{\prime }$ and $\overline{S}_0$ are the
zero-temperature values of corresponding parameters, $c=\sqrt{4J\gamma
_0(2J\gamma _0+J^{\prime }\gamma _0^{\prime })}$ is the spin-wave velocity.
The corresponding temperature dependences of $\gamma $ and $\gamma ^{\prime
} $ are given by 
\begin{eqnarray}
\gamma &=&\gamma _0-\frac{\pi ^2T^4}{120c^3\sqrt{JJ^{\prime }\gamma _0\gamma
_0^{\prime }}} \\
\gamma ^{\prime } &=&\gamma _0^{\prime }-\frac{\pi ^2T^4}{120c^3}\sqrt{\frac{%
J\gamma _0}{(J^{\prime }\gamma _0^{\prime })^3}}
\end{eqnarray}

In the case of small interlayer couplings $J^{\prime }/J\ll 1$ and higher
temperatures, logarithmic singularities occur, and we can pick out them from
the integrals in (\ref{gg1}), (\ref{MagnF}) and (\ref{pureAF}) in the same
way as discussed in Ref. \cite{Our1st}. In the quantum regime which takes
place at not too low temperatures, where 
\begin{eqnarray}
J^{\prime }S &\ll &T\ll JS\,\,\,\;\text{(FM)}  \nonumber \\
(JJ^{\prime })^{1/2}S &\ll &T\ll |J|S\,\,\,\text{(AFM)}  \label{QuReg}
\end{eqnarray}
we obtain 
\begin{mathletters}
\label{SlnQu}
\begin{eqnarray}
\overline{S} &=&S-\frac T{4\pi JS}\ln \frac T{J^{\prime }S}%
\;\;\;\;\;\;\;\;\;\;\;\text{(FM),} \\
\overline{S} &=&\overline{S}_0-\frac T{4\pi |J|\gamma }\ln \frac{T^2}{%
8JJ^{\prime }\gamma \gamma ^{\prime }}\,\;\;\text{(AFM),}
\end{eqnarray}
with $\gamma \simeq \gamma _0$ (the 2D values (\ref{D2}) can be used for $%
\gamma _0$ and $\overline{S}_0$) and $\gamma ^{\prime }$ being defined by
the equation 
\end{mathletters}
\begin{mathletters}
\label{Sln}
\begin{eqnarray}
\gamma ^{\prime } &=&S-\frac T{4\pi JS}\left( \ln \frac T{J^{\prime }\gamma
^{\prime }}-1\right) \;\;\;\;\;\;\;\;\text{(FM),} \\
\gamma ^{\prime } &=&\overline{S}_0-\frac T{4\pi |J|\gamma }\left( \ln \frac{%
T^2}{8JJ^{\prime }\gamma \gamma ^{\prime }}-1\right) \;\text{(AFM).}
\end{eqnarray}
so that $\gamma _0^{\prime }=\overline{S}_0.$ Note that in this case the
infrared cutoff for the integrals over quasimomenta is 
\end{mathletters}
\begin{equation}
q_0=\left\{ 
\begin{array}{cc}
(T/JS)^{1/2} & \text{(FM)} \\ 
T/c & \text{(AFM)}
\end{array}
\right.  \label{q0}
\end{equation}
($c=\sqrt{8}|J|\gamma $) rather than the boundary of the Brillouin zone.
Since $q_0\ll 1,$ the continuum approximation for the excitation spectrum
(and also interaction vertex) can be used in the quantum regime. Note that
owing to the thermodynamic identity $(\partial \overline{S}/\partial T)_{%
{\cal S}}=(\partial {\cal S}/\partial h)_T$ (with ${\cal S}$ being the
entropy, $h$ the magnetic field) the presence of $T\ln T$-terms in the
magnetization of a ferromagnet may be of interest in connection with the
adiabatic cooling (see, e.g., Ref. \cite{Vons}).

For the critical temperatures in the regime (\ref{QuReg}) we obtain from (%
\ref{SlnQu}) the results 
\begin{eqnarray}
T_C &=&\frac{4\pi JS^2}{\ln (T/J^{\prime }\gamma _c^{\prime })}\,,
\label{TMqu} \\
\,\,T_N &=&\frac{4\pi |J|\gamma _c\overline{S}_0}{\ln (T^2/8JJ^{\prime
}\gamma _c\gamma _c^{\prime })}  \nonumber
\end{eqnarray}
with $\gamma _c=\gamma (T_M)\simeq \gamma _0$ and $\gamma _c^{\prime
}=\gamma ^{\prime }(T_M)=T_M/4\pi |J|\gamma $. Comparing these results with
the criteria of quantum regime (\ref{QuReg}) we obtain the condition of
applicability of the results (\ref{TMqu}) as $2\pi S\ll \ln (J/J^{\prime }).$
It is important that $\gamma _c^{\prime }\ll \gamma ^{\prime }$ and the
interlayer coupling is strongly renormalized with the temperature. At the
same time, only ground-state (quantum) renormalizations are important for
the intralayer coupling at $|J^{\prime }|\ll |J|.$

In the case of large $S$ (again supposing $T\gg |J|S$) we obtain for both
ferro- and antiferromagnet 
\begin{equation}
\overline{S}=S-\frac T{4\pi |J|S}\ln \frac{32JS}{J^{\prime }\gamma ^{\prime }%
}  \label{SlnCl}
\end{equation}
with 
\begin{equation}
\gamma ^{\prime }=S-\frac T{4\pi |\,J\,|S}\left( \ln \frac{32JS}{J^{\prime
}\gamma ^{\prime }}-1\right)
\end{equation}
This leads to the expression for the critical temperature of a classical
magnet with $1\ll \ln (J/J^{\prime })\ll 2\pi \,S$ 
\begin{equation}
T_M=\frac{4\pi |J|S^2}{\ln (32JS/J^{\prime }\gamma _c^{\prime })}.
\label{TMcl}
\end{equation}
where $\gamma _c^{\prime }=T_M/4\pi |J|S.\;$As it should be, the critical
temperature is the same for the classical ferro- and antiferromagnetic case.
With the logarithmic accuracy we reproduce in this case the well-known
results where $\gamma _c^{\prime }/S\rightarrow 1$ (see, e.g., Ref. \cite
{Joungh}). Note that the factor of $32$ which is often neglected leads to
significant lowering of $T_M$ as well as above-considered temperature
dependence of $\gamma ^{\prime }$.

\subsection{Mean-field Schwinger-boson approach.}

\label{SB}Similar results can be obtained within the Schwinger-boson
representation. This is performed in the same way as in Refs.\cite
{Arovas,Yosh,Sarker}. The Heisenberg Hamiltonian is written down in the form 
\begin{eqnarray}
H &=&-\frac 12\sum_{ij}J_{ij}\left[ \frac 14(s_{i\uparrow }^{\dagger
}s_{i\uparrow }-s_{i\downarrow }^{\dagger }s_{i\downarrow })(s_{j\uparrow
}^{\dagger }s_{j\uparrow }-s_{j\downarrow }^{\dagger }s_{j\downarrow
})+s_{i\uparrow }^{\dagger }s_{i\downarrow }s_{j\downarrow }^{\dagger
}s_{j\uparrow }\right]  \nonumber \\
&&\ \ \ \ -\mu \sum_i(s_{i\uparrow }^{\dagger }s_{i\uparrow
}+s{}_{i\downarrow }^{\dagger }s_{i\downarrow })  \label{HSB}
\end{eqnarray}
where the chemical potential of bosons is introduced to take into account
the constraint (\ref{ccond}).

In the ferromagnetic case we subtract from the Hamiltonian (\ref{HSB}) the
term 
\begin{equation}
H_c=\frac 18\sum_{ij}J_{ij}(s_{i\uparrow }^{\dagger }s_{i\uparrow
}{}+s_{i\downarrow }^{\dagger }s_{i\downarrow })(s_{j\uparrow }^{\dagger
}s_{j\uparrow }{}+s_{j\downarrow }^{\dagger }s_{j\downarrow })\equiv \frac{%
J_0S^2}2  \label{Hc}
\end{equation}
to obtain 
\begin{equation}
\widetilde{H}=-\frac 14\sum_{<ij>}J_{ij}:{\cal F}_{ij}^{\dagger }{\cal F}%
_{ij}:-\mu \sum_i(s_{i\uparrow }^{\dagger }s_{i\uparrow }{}+s_{i\downarrow
}^{\dagger }s_{i\downarrow })  \label{SwHF}
\end{equation}
where $\widetilde{H}=H-H_c,$ ${\cal F}_{ij}=\sum_\sigma s_{i\sigma
}^{\dagger }s_{j\sigma },\,$and $:...:$ stands for the normal ordering$.$
Further the tilde at the Hamiltonian $H$ will be dropped. Introducing the
averages of the Bose operators 
\begin{equation}
\gamma _{ij}=\langle {\cal F}_{ij}\rangle =\langle {\cal F}_{ij}^{\dagger
}\rangle
\end{equation}
we derive the mean-field Hamiltonian 
\begin{equation}
H_{MF}=-\frac 12\sum_{<ij>}J_{ij}\gamma _{ij}{\cal F}_{ij}-\mu
\sum_i(s_{i\uparrow }^{\dagger }s_{i\uparrow }{}+s_{i\downarrow }^{\dagger
}s_{i\downarrow })
\end{equation}
Such a procedure can be justified if we generalize the Schwinger-boson
representation to the $SU(N)$ model with arbitrary $N$ by introducing the
operators $s_{im}^{\dagger }$ ($m=1...N$) and consider the limit $%
N\rightarrow \infty $ \cite{Arovas}.

In the quasi-2D case there are only two independent values of $\gamma _{ij}:$%
\begin{equation}
\gamma _{ij}=\left\{ 
\begin{array}{cc}
\gamma & i,j\,\text{within the same plane} \\ 
\gamma ^{\prime } & \text{otherwise}
\end{array}
\right.
\end{equation}
Introducing $\lambda =-\mu -\gamma J_0$ and passing to quasimomentum
representation we obtain 
\begin{equation}
H_{MF}=\sum_{{\bf q}\sigma }E_{{\bf q}}s_{{\bf q}\sigma }^{\dagger }s_{{\bf q%
}\sigma }
\end{equation}
where $E_{{\bf q}}=\lambda -\Gamma _{{\bf q}}.$ Note that in the absence of
external magnetic field the spectrum of bosons is doubly degenerate. The
self-consistent equations have the form 
\begin{eqnarray}
\gamma &=&\sum_{{\bf k}\sigma }N_{{\bf k}\sigma }\cos k_x,\,\,\gamma
^{\prime }=\sum_{{\bf k}\sigma }N_{{\bf k}\sigma }\cos k_z  \nonumber \\
2S &=&\sum_{{\bf k}\sigma }N_{{\bf k}\sigma }
\end{eqnarray}
As well as in Refs.\cite{Yosh,Sarker}, at low enough temperatures the Bose
condensation takes place. Introducing external magnetic field (see Sect. \ref
{Field}) removes the degeneracy of the boson spectrum, and only one of two
bosons is condensed. Let $N_{{\bf k}\uparrow }$ (but not $N_{{\bf k}%
\downarrow })$ contain the condensate contribution at $k\rightarrow 0$: 
\begin{equation}
N_{{\bf k\uparrow }}\rightarrow N_{{\bf k}}+2n_B\delta _{{\bf k}0}
\label{CondF}
\end{equation}
where $2n_B$ is the density of condensed bosons. Thus the self-consistent
equations takes the same form as in the BKJ representation with $\overline{S}%
\rightarrow n_B,$ $N(E_f)=0.$

In the antiferromagnetic case we subtract the term (\ref{Hc})\ from the
Hamiltonian to obtain (cf. \cite{Arovas}) 
\begin{equation}
\widetilde{H}=-\frac 12\sum_{<ij>}J_{ij}:{\cal A}_{ij}^{\dagger }{\cal A}%
_{ij}:-\mu \sum_i(s_{i\uparrow }^{\dagger }s_{i\uparrow }{}+s_{i\downarrow
}^{\dagger }s_{i\downarrow })
\end{equation}
where ${\cal A}_{ij}=s_{i\uparrow }^{\dagger }s_{j\downarrow }^{\dagger }.$
Passing to the mean-field approximation we have 
\begin{equation}
H_{MF}=-\frac 12\sum_{<ij>}\gamma _{ij}J_{ij}({\cal A}_{ij}+{\cal A}%
_{ij}^{\dagger })-\mu \sum_i(s_{i\uparrow }^{\dagger }s_{i\uparrow
}{}+s_{i\downarrow }^{\dagger }s_{i\downarrow })
\end{equation}
where 
\begin{equation}
\gamma _{ij}=\gamma _{ij}=\langle {\cal A}_{ij}\rangle =\langle {\cal A}%
_{ij}^{\dagger }\rangle
\end{equation}
Diagonalizing the Hamiltonian obtained one obtains 
\begin{equation}
H_{MF}=\sum_{{\bf q}}E_{{\bf q}}(\alpha _{{\bf q}}^{\dagger }\alpha _{{\bf q}%
}+\beta _{{\bf q}}^{\dagger }\beta _{{\bf q}})
\end{equation}
where $E_{{\bf q}}=(\lambda ^2-\Gamma _{{\bf q}}^2)^{1/2}.$ Thus the
self-consistent equations take the form 
\begin{mathletters}
\label{Sl}
\begin{eqnarray}
\gamma &=&\sum_{{\bf k}}\frac{\Gamma _{{\bf k}}}{2E_{{\bf k}}}\cos k_x(N_{%
{\bf k}\uparrow }{}+N_{{\bf k\downarrow }}+1), \\
\,\,\gamma ^{\prime } &=&\sum_{{\bf k}}\frac{\Gamma _{{\bf k}}}{2E_{{\bf k}}}%
\cos k_z(N_{{\bf k}\uparrow }{}+N_{{\bf k\downarrow }}+1) \\
2S &=&\sum_{{\bf k}}\frac{\Gamma _{{\bf k}}}{E_{{\bf k}}}(N_{{\bf k}\uparrow
}{}+N_{{\bf k\downarrow }}+1)-1
\end{eqnarray}
As well as in the ferromagnetic case, only $N_{{\bf k}\uparrow }$ contains
the condensate contribution. Picking out this as 
\end{mathletters}
\begin{equation}
N_{{\bf k\uparrow }}/E_{{\bf k\uparrow }}\rightarrow N_{{\bf k}}/E_{{\bf k}%
}+n_B(\delta _{{\bf k}0}+\delta _{{\bf kQ}})
\end{equation}
(${\bf Q}=(\pi ,\pi ,\pi )$ is the wavevector of the antiferromagnetic
structure) we get the SSWT\ equations (\ref{EqIAF}) with $\coth (E_f/T)=1,$ $%
\overline{S}\rightarrow n_B.$

The corrections to above results can be obtained within the $1/N$-expansion
in a generalized Heisenberg $SU(N)$ model (see, e.g., Refs. \cite
{Arovas,ArovasBook,Starykh,Starykh1,OurCP}). As argued in the introduction
(see also Sect.\ref{Interact}), the same results can be more easily obtained
by higher-order $1/S$-expansion. Thus the BKJ approach turns out to be more
practical than the Schwinger-boson one.

\section{SSWT of the easy-axis 2D magnets}

\label{E-ax}Consider now the 2D magnets with the easy-axis anisotropy.
Besides the spin-wave excitations, the topological excitations (domain
walls) contribute to thermodynamic quantities (see, e.g., discussion in Ref. 
\cite{Levanjuk}). Such excitations cannot be taken into account in the
approach under consideration. However, in the limit of small anisotropy 
\begin{equation}
D/|J|\ll 1,\eta \ll 1  \label{smalla}
\end{equation}
one can expect that the non-spin-wave excitations are important only in a
narrow critical region. Outside this region thermodynamics can be described
in terms of spin waves. Thus we restict ourselves to the case where (\ref
{smalla}) is satisfied.

Consider first the ferromagnetic case. Decoupling four-fold terms in the
Hamiltonian we obtain 
\begin{equation}
H=\sum_{{\bf k}}E_{{\bf k}}b_{{\bf k}}^{\dagger }b_{{\bf k}}{}+E_f\sum_{{\bf %
k}}c_{{\bf k}}^{\dagger }c_{{\bf k}}
\end{equation}
where 
\begin{eqnarray}
E_{{\bf k}} &=&\lambda -\Gamma _{{\bf k}},\,\,\,E_f=(2S+1)\lambda
\label{Eka} \\
\lambda &=&J_0(\gamma +\eta \overline{S})+D\left[ (2S-1)-4\langle
b_i^{\dagger }b_i\rangle \right] -\mu  \nonumber \\
\Gamma _{{\bf k}} &=&J_{{\bf k}}\left[ \gamma +\eta \langle b_i^{\dagger
}b_{i+\delta }\rangle \right]  \nonumber
\end{eqnarray}
and 
\begin{equation}
\gamma =\overline{S}+\langle b_i^{\dagger }b_{i+\delta }\rangle  \label{ksi1}
\end{equation}
It should be noted that the expression for the excitation spectrum (\ref{Eka}%
) is in fact the first-order $1/S$ expansion result. In particular, the
spectrum (\ref{Eka})\ violates the requirement of vanishing of single-site
anisotropy at $S=1/2$ (this situation is discussed in Ref. \cite{Rast2}). To
correct this inconsistency we perform two replacements in the spectrum (\ref
{Eka}), which can be justified by calculating higher-order terms in $1/S$: 
\begin{eqnarray}
(2S-1)-4\langle b_i^{\dagger }b_i\rangle &\rightarrow &(2S-1)\left[
1-2\langle b_i^{\dagger }b_i\rangle /S\right] \rightarrow (2S-1)(\overline{S}%
/S)^2 \\
\overline{S}-\langle b_i^{\dagger }b_{i+\delta }\rangle &\rightarrow
&S\left[ 1-2\langle b_i^{\dagger }b_i\rangle /S\right] \left[ 1+\langle
b_i^{\dagger }b_i\rangle /S-\langle b_i^{\dagger }b_{i+\delta }\rangle
/S\right] \rightarrow \overline{S}^2/\gamma  \nonumber
\end{eqnarray}
Then the boson spectrum takes the form 
\begin{equation}
E_{{\bf k}}=\gamma (J_0-J_{{\bf k}})+JS\Delta -\mu ,  \label{EkAnis}
\end{equation}
where 
\begin{equation}
\Delta (T)=\left[ (2S-1)D/|JS|+(J_0S/J\gamma )\eta \right] (\overline{S}/S)^2
\label{Del}
\end{equation}
is the dimensionless energy gap renormalized by spin-wave interactions. The
system of the self-consistent equations reads 
\begin{eqnarray}
\gamma &=&\overline{S}+\frac 1{J_0}\sum_{{\bf k}}J_{{\bf k}}N_{{\bf k}}
\label{pure} \\
\overline{S} &=&S-\sum_{{\bf k}}N_{{\bf k}}+(2S+1)N(E_f)  \nonumber
\end{eqnarray}

In the antiferromagnetic case we obtain 
\begin{equation}
H=\sum_{{\bf k}}E_{{\bf k}}(\alpha _{{\bf k}}^{\dagger }\alpha _{{\bf k}%
}+\beta _{{\bf k}}^{\dagger }\beta _{{\bf k}})+E_f\sum_{{\bf k}}(c_{{\bf k}%
}^{\dagger }c_{{\bf k}}+d_{{\bf k}}^{\dagger }d_{{\bf k}})
\end{equation}
with the spectrum 
\begin{eqnarray}
E_{{\bf k}} &=&\sqrt{\lambda ^2-\Gamma _{{\bf k}}^2},\;E_f=(2S+1)\lambda \\
\lambda &=&\gamma J_0+|JS|\Delta -\mu ,\;\Gamma _{{\bf k}}=\gamma J_{{\bf k}}
\nonumber
\end{eqnarray}
and $\Delta $ is the same as in (\ref{Del}). The system of the
self-consistent equations takes the form 
\begin{eqnarray}
\gamma &=&\overline{S}+\sum_{{\bf k}}\frac{\Gamma _{{\bf k}}}{2E_{{\bf k}}}%
\cos k_x\coth \frac{E_{{\bf k}}}{2T} \\
\overline{S} &=&(S+\frac 12)\coth \frac{E_f}{2T}-\sum_{{\bf k}}\frac \lambda
{2E_{{\bf k}}}\coth \frac{E_{{\bf k}}}{2T}
\end{eqnarray}
Note that the proportionality of the gap in the spin-wave spectrum to the
squared sublattice magnetization was obtained earlier within the
renormalized spin-wave theory \cite{Rast2,Oguchi}, which takes into account
the influence of spin-wave interactions on the spectrum in a
non-self-consistent way, and it is in agreement with the experimental data 
\cite{Nagata}.

Using the smallness of anisotropy and picking out the logarithmic
singularities in the same way as in Sect.\ref{TempDep} we obtain 
\begin{eqnarray}
\overline{S} &=&S-\frac T{4\pi JS}\ln \frac T{JS\Delta }\,\,,\,\,\;\,\,\,\,%
\,\,\,\,\,\,\,\,\,\,\,\,\,\text{(}\,\text{FM),}  \label{Sa} \\
\overline{S} &=&\overline{S}_0-\frac T{4\pi |J|\gamma }\ln \frac{T^2}{%
8(J\gamma )^2\Delta }\,,\,\,\,\,\;\text{(AFM).}  \nonumber
\end{eqnarray}
Unlike the quasi-2D case, we have the unphysical result $\Delta (T_M)=0$
because of the proportionality of the gap to $(\overline{S}/S)^2$ (in fact a
finite value of the gap at $T=T_M$ should be caused by topological effects
which are not taken into account). Thus we are unable to describe the
dependence $\Delta (T)$ close to $T_M.$ Denoting $\Delta _c=\Delta (T_M)$ we
have for the critical temperature at $2\pi S\ll \ln (1/\Delta )$%
\begin{eqnarray}
T_C &=&\frac{4\pi JS^2}{\ln (T/JS\Delta _c)}\,,\,  \label{TMaqu} \\
\,\,T_N &=&\frac{4\pi |J|\overline{S}_0\gamma _c}{\ln [T^2/8(J\gamma
_c)^2\Delta _c]}  \nonumber
\end{eqnarray}
In the case of large $S$ we obtain for both ferro- and antiferromagnets 
\begin{equation}
\overline{S}=S-\frac T{4\pi |J|S}\ln \frac{32}\Delta  \label{Sacl}
\end{equation}
This leads to the expression for the critical temperature of a classical
magnet with $1\ll \ln (1/\Delta )\ll 2\pi \,S$ 
\begin{equation}
T_M=\frac{4\pi |J|S^2}{\ln (32/\Delta _c)}.  \label{TMacl}
\end{equation}
To leading logarithmic accuracy we can put $\Delta _c=\Delta (0)$ in the
above results$.$ More correct calculation of $\Delta _c,$ as well as the
corrections to the results (\ref{TMaqu}) and (\ref{TMacl}) will be obtained
in Sect. \ref{Interact}. Note also that in the approximation $\Delta
(T)=\Delta (0),$ i.e. at neglecting the temperature dependence of the gap,
we reproduce correctly the mean field result in the Ising limit, 
\begin{equation}
\overline{S}=SB_S(J_0S\overline{S}/T)
\end{equation}
where $B_S(x)$ is the spin-$S$ Brillouin function.

\section{Influence of the external magnetic field and the magnetic
susceptibility}

\label{Field}In this section we consider the influence of a weak external
magnetic field $h$ in a ferromagnet. This is described by the additional
term in Hamiltonian, 
\begin{equation}
H_h=-h\sum_iS_i^z.
\end{equation}
The magnetic field results in an increase of magnetization, so that the
total magnetization can be represented as 
\begin{equation}
\overline{S}=\overline{S}_{\text{sp}}+\overline{S}_{\text{ind}}
\label{SpInd}
\end{equation}
where $\overline{S}_{\text{sp}}=\overline{S}(h=0)$ is the spontaneous
magnetization, $\overline{S}_{\text{ind}}$ is the field-induced part. Owing
to the second term in (\ref{SpInd}) the temperature dependence $\overline{S}%
(T)$ is changed: sharply decreasing in the vicinity of $T_M$, $\overline{S}$
nevertheless vanishes only in the limit $T\rightarrow \infty $ We consider a
possible approach to the description of such a behavior in both versions of
SSWT that are based on the Dyson-Maleev representation (or its
generalization with the use of the BKJ\ representation) and Schwinger-boson
representation.

First we use the BKJ representation. The calculations, that are similar to
described above, result in the equations (\ref{gg1}) and (\ref{MagnF}) with
the spectrum of spin waves 
\begin{equation}
E_{{\bf k}}=\gamma (J_0-J_{{\bf k}})+h-\mu _0  \label{Ekh}
\end{equation}
where $\mu _0$ is the chemical potential in the absence of magnetic field: $%
\mu _0=0$ at $T<T_C$ and $\mu _0(T>T_C)$ is determined from the condition $%
\overline{S}(T,h=0)=0$.

Formally, the spectrum (\ref{Ekh})\ has the same form as in the case of an
anisotropic magnet (\ref{EkAnis}) (we can associate with the anisotropy the
effective ``magnetic field'' $h_A=JS\Delta $). However, there is an
important difference: in the case of the ``true'' magnetic field the
chemical potential is taken at $h=0,$ so that the phase transition with
vanishing $\overline{S}$ is absent (see below), while in the case of
anisotropic magnet it should be determined in the presence of anisotropy
field $h_A,$ and $\overline{S}$ vanishes at $T_C$. However, at $T\ll T_C$
this difference is not important ($\mu =\mu _0=0$ in this region) and the
magnetic anisotropy can be also described by introducing the
temperature-dependent magnetic anisotropy field $h_A.$

The temperature dependence of magnetization obtained by numerical solution
of Eqs.(\ref{gg1}) and (\ref{MagnF}) with the spectrum (\ref{Ekh})\ is shown
in Fig.\ref{FigField}. At low temperatures $T\ll T_C$ we have $\overline{S}_{%
\text{sp}}\gg \overline{S}_{\text{ind}}$ and magnetization has mainly
exchange origin. On the other hand, at $T>T_C$ magnetization is entirely
caused by influence of an external field and 
\begin{equation}
\overline{S}\simeq \chi _0^{zz}h,\;T\gg T_C
\end{equation}
where 
\begin{equation}
\chi _0^{zz}=\left( \frac{\partial \overline{S}}{\partial h}\right) _{h=0}=%
\frac 1{4T}\sum_{{\bf q}}\frac 1{\sinh ^2(E_{{\bf q}}/2T)}-\frac{(S+1/2)^2}{%
T\sinh ^2(E_f/2T)}  \label{Hi00}
\end{equation}
The first term in (\ref{Hi00}) differs from the result of the spin-wave
theory by the form of the spectrum only, and the last term describes the
correction owing to the kinematical interaction. In a narrow region near $%
T_C $ both contributions in (\ref{SpInd}) are of the same order, and
magnetization considerably differs from its zero-field value.

It follows from (\ref{Hi00}) that $\chi _0^{zz}\propto (T-T_M)^{-2}$ so that
the critical exponent is $\gamma =2.$ Note that magnon-magnon interactions
are taken into account in (\ref{Hi00}) only by renormalization of
single-particle spectrum. It is possible to improve result (\ref{Hi00}) by
taking into account two-particle interactions in a RPA-type way, i.e. by
considering the sum of the one-loop diagrams. This is performed in the next
section.

Now we consider the influence of external field in the Schwinger-boson
representation. Carrying out the calculations similar to Sect.\ref{SB} we
find for the boson spectrum 
\begin{equation}
E_{{\bf k}\sigma }=\gamma _\sigma (J_0-J_{{\bf k}})-\frac 12h\sigma -\mu
\end{equation}
where $\gamma _\sigma =\langle s_{i\sigma }^{\dagger }s_{i\sigma }\rangle $.
The expression for the magnetization has the form 
\begin{equation}
\overline{S}=\frac 12\sum_{{\bf k}}(N_{{\bf k\uparrow }}-N_{{\bf k\downarrow 
}})+n_B  \label{Szh}
\end{equation}
where we have taken into account the possibility of condensation of bosons
with up ``spins''. There is also the condition of spin conservation at each
site 
\begin{equation}
S=\frac 12\sum_{{\bf k}}(N_{{\bf k\uparrow }}{}+N_{{\bf k\downarrow }})+n_B
\label{Constrh}
\end{equation}
At not too high temperatures $T<T_h$, where $T_h$ is determined by the
conditions 
\begin{equation}
S=\frac 12\sum_{{\bf k}}(N_{{\bf k\uparrow }}+N_{{\bf k\downarrow }}),\;\mu
=-h/2,
\end{equation}
the branch $E_{{\bf k\uparrow }}$ is gapless and $n_B>0$. At $T>T_h$ both
branches have a gap, and the condition (\ref{Constrh}) with $n_B$ $=0$
determines the common chemical potential. Thus the Schwinger-boson
representation also allows us to describe the behavior of magnetization in
the whole field interval, and the expression (\ref{Szh}) just describes
magnetization as a sum of spontaneous and field-induced components.

Up to now we have considered the small magnetic field values $h\ll J.$ In
the opposite limit one can neglect the dispersion of boson spectrum (\ref
{Ekh}) and derive by using the BKJ representation the standard result 
\begin{equation}
\overline{S}=SB_S(Sh/T)\text{.}  \label{Brillouen}
\end{equation}
It should be noted that the correct result (\ref{Brillouen}) is obtained
only due to presence of pseudofermions, the Bose field alone leading to the
unphysical phase transition with vanishing of magnetization at $T\sim h.$

A somewhat different situation takes place in the Schwinger boson
representation. In the case $h\gg J$ we have $n_B\equiv 0$ and the equation
for $x=\exp (-\mu /T)$ has the form 
\begin{equation}
\frac{x\cosh (h/2T)-1}{x^2+1-2x\cosh (h/2T)}=S  \label{EqLH}
\end{equation}
The solution to this equation reads 
\begin{equation}
x=(1+\frac 1{2S})\cosh \frac h{2T}+\frac 1{2S}\sqrt{(2S+1)^2\cosh ^2\frac h{%
2T}-4S(S+1)}
\end{equation}
With the use of (\ref{EqLH}) we obtain the expression for the magnetization 
\begin{equation}
\overline{S}=S\frac{x\sinh (h/2T)}{x\cosh (h/2T)-1}\simeq S\tanh
(h/2T)\equiv SB_{1/2}(h/2T)
\end{equation}
Thus in the limit of large magnetic fields the Schwinger-boson approach
reproduces correct results only for $S=1/2.$

\section{Fluctuation corrections to SSWT for 2D and quasi-2D magnets}

\label{Interact}As already discussed, SSWT overestimates the value of $T_M.$
In particular, for the simple cubic lattice the SSWT result for $%
S\rightarrow \infty $ is $T_M/S^2=1.803|J|$. At the same time, the result of
the spherical model (see, e.g., \cite{Baxter,Nag}) in this lmit reads 
\begin{equation}
\frac{S^2}{3T_M}=\sum_{{\bf k}}\frac 1{J_0-J_{{\bf k}}}  \label{MSph}
\end{equation}
which coincides with the corresponding result of the Tyablikov approximation 
\cite{Tyab}. One obtains from (\ref{MSph}) $T_M/S^2=1.319|J|$ which is close
to the result of the high-temperature series expansion (see, e.g., Refs. 
\cite{Joungh,Tyab}). As pointed in Sect. \ref{Quasi-2D}, for $S=1/2$ the
value of $T_M$ is overestimated by $1.2$ times.

In the quasi-2D case the formulas (\ref{TMqu}) and (\ref{TMcl}) (and the
corresponding results of 2D case with small easy-axis anisotropy (\ref{TMaqu}%
) and (\ref{TMacl})) coincide with the result of the Tyablikov approximation
to logarithmic accuracy and thus seem to be correct. However, this accuracy
is also insufiicient to treat experimental data (see detailed discussion in
Ref. \cite{Our1/N}) and the overestimation of $T_M$ reaches $1.7\div 2.0$
times for the quasi-2D case and nearly $1.5$ times for anisotropic 2D case
(the reason of weaker overestimation of $T_M$ in the anisotropic case will
be explained below). Thus in the quasi-2D magnets and 2D magnets with small
easy axis anisotropy (in both cases $T_M\ll |J|S^2$) the overestimation of $%
T_M$ even higher then in $3D$ case.

The values of critical exponents derived above ($\beta =\nu =1$ and $\gamma
=2$) are also in drastic discrepance with the molecular-field values ($\nu
=\beta =1/2,$ $\gamma =1$), experimental data ($\nu =0.7,$ $\beta =0.33,$ $%
\gamma =1.4$) for isotropic magnets and exact values ($\nu =1,$ $\beta =1/8,$
$\gamma =7/4$) for easy-axis magnets, which are known from the Onsager
solution of 2D\ Ising model. Thus SSWT describes poorly the critical
behavior.

At the same time, SSWT describes much better local properties (e.g., the
pair spin correlation function at neighbor sites) than those determined by
the scale of the correlation length. Indeed, at $S=1/2$ the Tyablikov
approximation yields the unphysical result $\Delta {\cal E}(T_C)={\cal E}%
(T_C)-{\cal E}(0)<0$ \cite{LiuSH}. In the limit $S\rightarrow \infty $ this
approximation gives $\Delta {\cal E}(T_C)/|{\cal E}(0)|=0.6$ which is also
lower than the value which can be derived from the calculations in Sect.\ref
{Quasi-2D} ($0.84$). Besides that, the Tyablikov approximation implies a not
quite correct form of the excitation spectrum at low temperatures. In
particular, the spin-wave stiffness demonstrates the $T^{3/2}$ dependence at
low temperatures, instead of $T^{5/2}$ one.

Generally speaking, the properties on the scales of order of correlation
length cannot be treated correctly within one-particle picture, and the
Tyablikov approximation gives only rough (but rather successful) description
of these. A regular way of describing thermodynamics at not too low
temperatures within spin-wave theory is to consider collective excitations
rather than one-particle ones. For low-dimensional magnets with $T_M\ll
|J|S^2,$ where large logarithms occur (see Sects. \ref{Q2dBKJ} and \ref{E-ax}%
) and fluctuations have 2D nature in a broad temperature region (except for
the critical region), this can be performed analytically in a close analogy
with the isotropic magnets of the dimensionality $d=2+\varepsilon $ (where $%
\beta =1+{\cal O}(\varepsilon )$, see, e.g., Ref. \cite{Brezin}).

In this Section we take into account the interaction corrections to the SSWT
results for the magnets with small interlayer coupling and/or anisotropy.
Consider first the 2D Heisenberg magnet with the easy-axis anisotropy. In
the ferromagnetic case we have 
\begin{equation}
H=\sum_{{\bf q}}E_{{\bf q}}^0b_{{\bf q}}^{\dagger }b_{{\bf q}}+\frac 14\sum_{%
{\bf q}_1...{\bf q}_4}\varphi ({\bf q}_1,{\bf q}_2{\bf ;q}_3,{\bf q}_4)b_{%
{\bf q}_1}^{\dagger }b_{{\bf q}_2}^{\dagger }b_{{\bf q}_3}b_{{\bf q}%
_4}\delta _{{\bf q}_1+{\bf q}_2,{\bf q}_3+{\bf q}_4}  \label{HInt}
\end{equation}
where 
\begin{eqnarray}
E_{{\bf q}}^0 &=&S(J_0-J_{{\bf q}})+|J|Sf  \nonumber \\
\varphi ({\bf q}_1,{\bf q}_2{\bf ;q}_3,{\bf q}_4) &=&J_{{\bf q}_3}+J_{{\bf q}%
_4}-J_{{\bf q}_1{\bf -q}_3}-J_{{\bf q}_1{\bf -q}_4}\simeq -2|J|({\bf q}_1%
{\bf q}_2+f)
\end{eqnarray}
and 
\begin{equation}
f=(2S-1)D/|JS|+(J_0/J)\eta S  \label{f}
\end{equation}
is the bare gap in the excitation spectrum. In the antiferromagnetic case,
we use the operators $B_{{\bf q}}$ which are Fourier transformation of $B_i$
of Eq. (\ref{BC}) and satisfy 
\begin{eqnarray}
a_{{\bf q}} &=&(B_{{\bf q}}+B_{{\bf q+Q}})/2  \nonumber \\
b_{-{\bf q}}^{\dagger } &=&(B_{{\bf q}}-B_{{\bf q+Q}})/2
\end{eqnarray}
where ${\bf Q}=(\pi ,\pi ,\pi )$ is the wavevector of the AFM structure.
Then, up to some unimportant constant, we have the Hamiltonian of the same
form (\ref{HInt}), but for the operators $B_{{\bf q}}$. Note that in this
case $E_{{\bf q}}^0$ in (\ref{HInt}) has not the meaning of an excitation
spectrum because of non-Bose commutation relations for $B_{{\bf q}}$: 
\begin{equation}
\lbrack B_{{\bf q}},B_{{\bf p}}^{\dagger }]=\delta _{{\bf q,p}}+\delta _{%
{\bf q,p+Q}}  \label{crel}
\end{equation}

The diagrams which give the first-order renormalizations of $E_{{\bf q}}$
and correspond to SSWT are shown in Fig. \ref{FigFluct}a (see, e.g., Ref. 
\cite{BKJBook} for the detailed description of this diagram technique).
Further on we suppose that all such renormalizations (which result in the
replacements $J\rightarrow J\gamma /S$ and $f\rightarrow \Delta $ in $E_{%
{\bf q}}^0$) are already performed and such diagrams can be omitted.

To obtain the corrections to SSWT, higher-order diagrams should be
considered. They lead to renormalization of one-particle energy (and
occurrence of the damping) and also to vertex corrections. As discussed
above, SSWT treats the excitation spectrum satisfactorily (this spectrum is
already renormalized by first-order diagrams). The calculations of damping
of spin-waves, which occurs only in the second order of perturbation theory,
shows that it is small in a broad temperature region\cite{Damp}. Thus only
vertex corrections should be taken into account. At not too low temperatures
($T\gg |J|S\Delta $) the RPA-type diagrams of Fig. \ref{FigFluct}b are most
important since each loop contains a logarithmic divergence of the type $\ln
(1/\Delta ).$ The integral equation for the vertex reads 
\begin{eqnarray}
\Phi ({\bf k,k-q;p-q,p}) &=&\varphi ({\bf k,k-q;p-q,p})  \label{EqVertex} \\
&&\ \ \ \ \ -\frac T{(J\gamma )^2}\sum_{{\bf s}}\frac{\varphi ({\bf %
k,k-q;s-q,s)}}{(s^2+\Delta )[({\bf s-q})^2+\Delta ]}\Phi ({\bf s,s-q;p-q,p)}
\nonumber
\end{eqnarray}
(we have retained only the contribution of the modes with the Matsubara
frequency $\omega _n=0,$ which yields the logarithmic divergence, and
dropped the terms with $\omega _n\neq 0$ with simultaneous cutting the
summation over quasimomenta\ at the wavevector $q_0$ which is determined by (%
\ref{q0})). As can be seen from (\ref{Sacl}), the account of logarithmically
divergent terms in the classical case can also be performed in the continuum
approximation with $q_0^2=32.$ The result of solution of the equation (\ref
{EqVertex}) (see Appendix B) in the 2D case reads 
\begin{eqnarray}
\Phi ({\bf k,k-q;p-q,p}) &=&\frac{2|J|\gamma {\bf k}({\bf q-p})}{\overline{S}%
_0-(T/2\pi |J|\gamma )\ln [q_0/\max (\Delta ^{1/2},q)]}  \nonumber \\
&&\ \ \ \ \ -2|J|f\left[ 1-\frac T{\pi |J|\gamma ^2}\frac{{\bf q}({\bf q+k-p)%
}}{q^2}\ln \frac q{\Delta ^{1/2}}\right]  \label{FAnsw}
\end{eqnarray}
Note that the logarithmic corrections to the vertex in the isotropic case
were obtained earlier in Ref. \cite{ChubF}. For the static (staggered)
non-uniform longitudinal susceptibility (for AFM case the shift ${\bf q}%
\rightarrow {\bf q}+{\bf Q}$ is to be performed) we obtain from the diagrams
of Fig.\ref{FigFluct}c the result 
\begin{eqnarray}
\chi _q^{zz} &=&\frac{\chi _{q0}^{zz}}{1+(|J|\gamma /2\overline{S})q^2\chi
_{q0}^{zz}}  \nonumber \\
\ &=&\frac{(\overline{S}/\overline{S}_0)\chi _{q0}^{zz}}{1-(T/2\pi |J|\gamma 
\overline{S}_0)\ln [q_0/\max (\Delta ^{1/2},q)]}  \label{HiFluct}
\end{eqnarray}
where 
\begin{eqnarray}
\chi _{q0}^{zz} &=&\frac T{(J\gamma )^2}\sum_{{\bf p}}\frac 1{(p^2+\Delta )[(%
{\bf p-q})^2+\Delta ]}  \nonumber \\
\ &\simeq &\left\{ 
\begin{array}{cc}
T/[2\pi (J\gamma q)^2]\ln (q^2/\Delta ),\; & q^2\gg \Delta \\ 
\chi _0=T/[4\pi (J\gamma )^2\Delta ], & q^2\ll \Delta
\end{array}
\right.  \label{Hiq0}
\end{eqnarray}
is the ``bare'' longitudinal susceptibility. Thus, as well as in RPA for
itinerant magnets \cite{Moria}, the spin susceptibility is enhanced by the
interaction. It follows from the result (\ref{HiFluct}) that the excitation
spectrum has different forms at small and large enough momenta: 
\begin{equation}
\chi _q^{zz}\simeq \left\{ 
\begin{array}{cc}
\chi _{q0}^{zz}, & |J|q^2\chi _{q0}^{zz}\ll \overline{S}/S \\ 
2\overline{S}/(|J|\gamma q^2), & |J|q^2\chi _{q0}^{zz}\gg \overline{S}/S
\end{array}
\right.  \label{HiCases}
\end{equation}
The first line corresponds to the standard spin-wave contribution (it is
also subdivided in two cases as given by (\ref{Hiq0})). The second line
corresponds to non-spin-wave regime: at $q^2\gg \Delta $ one can neglect the
anisotropy and $\chi _q^{zz}\propto 1/q^2$ is given, in particular, by the
spherical model \cite{Nag}, which treats the spin excitations in essentially
non-spin-wave way. Depending on the temperature value, three cases are
possible.

\noindent (i) low temperatures, $T\ll T_M\sim 2\pi |J|S^2/\ln (q_0^2/\Delta
).$ Then the second condition in (\ref{HiCases}) cannot be satisfied and
thus the excitations in the whole Brillouin zone have spin-wave nature.

\noindent (ii) intermediate temperatures, $(\overline{S}/S)/\ln
(q_0^2/\Delta )\ll T/2\pi |J|S^2\ll \overline{S}/S$ ($T$ is of the same
order as $T_M$)$.$ Then at small enough $q$ we still have $\chi
_q^{zz}\simeq \chi _{q0}^{zz},$ but the second condition in (\ref{HiCases})
holds for large enough $q$\ where $\Delta \exp (2\pi |J|\gamma \overline{S}%
/T)\ll q^2<q_0^2$.

\noindent (iii) critical region, $T/2\pi |J|S^2\gg \overline{S}/S\;$($%
1-T/T_M\ll 1$). In this regime the first condition in (\ref{HiCases}) is
satisfied only for $q^2\ll \Delta $ (hydrodynamic region) whereas at all
other $q$ the condition in the second line of Eq.(\ref{HiCases}) is
satisfied.

The corrections to relative (sublattice) magnetization $\overline{\sigma }%
\equiv \overline{S}/\overline{S}_0$ (see diagrams of Fig.\ref{FigFluct}d)
are given by 
\begin{equation}
\overline{\sigma }=1-\frac T{|J|\gamma \overline{S}_0}\sum_{{\bf k}}\frac 1{%
k^2+\Delta }+\frac{T^2}{2(|J|\gamma )^3\overline{S}_0}\sum_{{\bf kq}}\frac{%
\Phi ({\bf k,k-q;k-q,k})}{(k^2+\Delta )^2[({\bf k-q})^2+\Delta ]}
\end{equation}
Integration leads to the result 
\begin{equation}
\overline{\sigma }=1-\frac t2\left[ \ln \frac{q_0^2}{\Delta _0}+4\ln \frac 1{%
\max (\overline{\sigma },t)}-2(1-\overline{\sigma })+\Phi _{\text{a}}\left(
t/\overline{\sigma }\right) \right]  \label{Res1}
\end{equation}
where $t=T/(2\pi |J|\overline{S}_0\gamma _0)$. The function $\Phi _{\text{a}%
} $ takes into account the (unknown) non-singular contribution of non-RPA
diagrams. Again, we have three temperature regions described above. In the
region (i) only first term in the square brackets is to be taken into
account and the magnetization demonstrates the spin-wave behavior (\ref{Sa})
and (\ref{Sacl}) for quantum and classical cases respectively. In the region
(ii) all the terms, except for the last, are important, which leads to
significant modification of the dependence $\overline{S}(T).$ The function $%
\Phi _{\text{a}}$ in both regimes (i) and (ii) can be neglected and the
result (\ref{Res1}) completely describes the behavior of magnetization in
these two regimes. Finally, in the region (iii) the contribution of $\Phi _{%
\text{a}}$ is of the same order as other terms in the square brackets. It
should be noted that the factor of $4$ before the second term in the square
brackets is the sum of $2$ which arises from the temperature renormalization
of $\Delta ^{1/2}(T)\propto \Delta _0^{1/2}\max (\overline{\sigma },t),$ and
also a contribution of $2$ arises from the vertex renormalization. Thus one
can see that in the case of small anisotropy (the same situation takes place
for small interlayer coupling, see below) the contribution from the
renormalization of single-particle spectrum and interaction vertex are of
the same order, so that SSWT is insufficient even outside the critical
region.

For the Curie (Neel) temperatures we obtain the equations 
\begin{mathletters}
\label{Res1abc}
\begin{eqnarray}
T_C &=&4\pi JS^2\left[ \ln \frac{T_C}{JS\Delta _0}+4\ln \frac{4\pi JS^2}{T_C}%
+C_{\text{F}}\right] ^{-1}\,\,\,\,\,\,\,\,\,\,\,\,\,\;\;\;\text{(FM)}
\label{Res1a} \\
T_N &=&4\pi J\overline{S}_0\gamma _0\left[ \ln \frac{T_N^2}{c^2\Delta _0}%
+4\ln \frac{4\pi |J|\overline{S}_0\gamma _0}{T_N}+C_{\text{AF}}\right]
^{-1}\;\text{(AFM)}  \label{Res1b} \\
T_M &=&4\pi JS^2\left[ \ln \frac{32}{\Delta _0}+4\ln \frac{4\pi |J|S^2}{T_M}%
+C_{\text{cl}}\right] ^{-1}\;\;\;\;\;\;\;\;\;\text{(classical)}
\label{Res1c}
\end{eqnarray}
with the constants $C_{\text{F,AF,cl}}=$ $-2-4\ln 2+\Phi _{\text{a}}^{\text{%
F,AF,cl}}(\infty )$ which are still not determined within our approach.
However, it is important that all the logarithmic terms are included in (\ref
{Res1abc}) and $C$ give only a small contribution to above results. The gap $%
\Delta _c$ at the ordering temperature, which remained indeterminate in Sect.%
\ref{E-ax}, can be now estimated as $\Delta _c\propto t^2.$ The coefficient
of proportionality is of order of unity and influences the constants $C$
only.

In the isotropic quasi-2D case the infrared cutoff for integrals over the
Brillouin zone is $J^{\prime }/J$ rather than $\Delta .$ Then we obtain in
the same way 
\end{mathletters}
\begin{equation}
\overline{\sigma }=1-\frac t2\left[ \ln \left( q_0^2\frac{|\,J\,|\gamma _0}{%
|J^{\prime }|\gamma _0^{\prime }}\right) +3\ln \frac 1{\max (\overline{%
\sigma },t)}-2(1-\overline{\sigma })+\Phi _{\text{ic}}\left( t/\overline{%
\sigma }\right) \right]  \label{Res2}
\end{equation}
and 
\begin{mathletters}
\label{Res2abc}
\begin{eqnarray}
T_C &=&4\pi JS^2\left[ \ln \frac{T_C}{|J^{\prime }|S}+3\ln \frac{4\pi JS^2}{%
T_C}+C_{\text{F}}^{\prime }\right]
^{-1}\,\,\,\,\,\,\,\,\,\,\,\,\,\;\;\;\;\;\;\;\;\;\;\;\;\;\;\text{(FM)}
\label{Res2a} \\
T_N &=&4\pi |\,J\,|S\overline{S}_0\gamma _0\left[ \ln \frac{T_N^2}{%
8|JJ^{\prime }|\gamma _0\gamma _0^{\prime }}+3\ln \frac{4\pi |\,J\,|%
\overline{S}_0\gamma _0}{T_N}+C_{\text{AF}}^{\prime }\right] ^{-1}\;\text{%
(AFM)}  \label{Res2b} \\
T_M &=&4\pi JS^2\left[ \ln \frac{32}{\Delta _0}+3\ln \frac{4\pi JS^2}{T_M}%
+C_{\text{cl}}^{\prime }\right] ^{-1}\;\;\;\;\;\;\;\;\;\;\;\;\;\;\;\;\;\;\;\;%
\text{(classical)}  \label{Res2c}
\end{eqnarray}
In this case we have $\gamma ^{\prime }(T)\propto \gamma _0^{\prime }\max (%
\overline{\sigma },t)$ which leads to that the coefficient at the second
term in the square brackets is $3$ (instead of $4$ in the anisotropic case).
This is why the interaction corrections are weaker in the anisotropic case:
within SSWT the above-mentioned coefficient is $1$ in the quasi-2D case
(which is $3$ times smaller than the correct value) and $2$ in the
anisotropic case (only $2$ times smaller than the correct value). Note that
the results (\ref{Res2}), (\ref{Res2abc}) are valid for all the four
combinations of the signs of intra- and inter-plane exchange integrals (for
mixed combinations, FM and AFM denote the type of the in-plane ordering).

The same results (\ref{Res1})-(\ref{Res2abc}) were obtained within the RG
approach in Ref. \cite{OurRG}. (Note that different sign at the third term
of square brackets of (\ref{Res1}) and (\ref{Res2}) is the misprint of this
paper). Derivation of general expressions for the case where both interlayer
coupling and anisotropy are of the same order can be also found in Ref.\cite
{OurRG}.

With neglect of the functions $\Phi _{\text{a}}(x)$ and $\Phi _{\text{ic}%
}(x),$ Eqs.(\ref{Res1}) and (\ref{Res2}) still yield unphysical behavior
near $T_M.$ The point $T^{*},$ where the derivative $\partial \overline{S}%
/\partial T$ diverges, can be determined from the condition 
\end{mathletters}
\begin{equation}
\overline{\sigma }(t^{*})/t^{*}\simeq \left\{ 
\begin{array}{cc}
3/2 & \text{quasi-2D} \\ 
2 & \text{easy-axis 2D}
\end{array}
\right.  \label{InvC}
\end{equation}
which should be used together with (\ref{Res1}) or (\ref{Res2}); $t^{*}$ is
the value of $t$ corresponding to $T^{*}.$ The condition

The functions $\Phi _{\text{a}}(x)$ and $\Phi _{\text{ic}}(x)$ describe the
crossover from an isotropic 2D Heisenberg to 2D Ising and 3D Heisenberg
behavior respectively. As discussed above, these functions give considerable
contributions in the crossover region between regimes (ii) and (iii) and in
the critical regime (iii), where essentially non-spin-wave excitations
should be taken into account. An account of these functions results in
slight decreasing the temperature $T^{*}$ in comparison with that given by (%
\ref{InvC}), and $T^{*}$ becomes the temperature of a rapid decrease of $%
\overline{S}$ (in fact, the characteristic temperature of a crossover). For
a quantum antiferromagnet, the calculation of $\Phi _{\text{ic}}(x)$ can be
performed within the $1/N$ expansion in $O(N)$ model \cite{Our1/N}. For an
arbitrary $x=t/\overline{\sigma }$, the result of this calculation is very
cumbersome. In the critical region ($x\gg 1$) it provides the correct
critical behavior \cite{Our1/N} 
\begin{equation}
\overline{\sigma }^2=\left[ \frac{T_{\text{Neel}}}{4\pi |J|\overline{S}%
_0\gamma _0}\right] ^{1-\beta _3}\left[ \frac 1{1-A_0}\left( 1-\frac T{T_{%
\text{Neel}}}\right) \right] ^{2\beta _3}  \label{MagnCr1}
\end{equation}
with $A_0\simeq 0.9635$ and $\beta _3=(1-8/\pi ^2N)/2\simeq 0.36.$ The value
of $C_{\text{AF}}^{\prime }$ obtained by this expansion is very small, $C_{%
\text{AF}}^{\prime }=-0.0660.$ Other critical exponents can also be
calculated within the $1/N$ expansion in $O(N)$ model (see, e.g., Ref. \cite
{Ma}): 
\[
\nu _3=1-32/3\pi ^2N\simeq 0.64,\;\gamma _3=2(1-12/\pi ^2N)\simeq 1.21 
\]
(note that the scaling relations are slightly violated because of
approximate character of this expansion for $N=3$). Thus the results of the
spherical model for the critical exponents above $T_M$ (see, e.g., Ref. \cite
{Nag}) become radically improved. In particular, the fluctuations correct
the critical behavior of magnetic susceptibility.

For practical purposes, it is useful to have simple interpolation
expressions for the functions $\Phi (x),$ which enable one to describe the
crossover temperature region. Taking into account the closeness of $T^{*}$
to $T_M$ which is given by (\ref{Res1abc}) and (\ref{Res2abc}) and using (%
\ref{InvC}), we can write down the simplest expressions for $\Phi (x)$ in
the form: 
\begin{eqnarray}
\Phi _{\text{a}}^{\text{F,AF,cl}}(x) &=&\frac x{\sqrt{x^2+1}}(C_{\text{%
F,AF,cl}}-2+8\ln 2)  \nonumber \\
\Phi _{\text{ic}}^{\text{F,AF,cl}}(x) &=&\frac x{\sqrt{x^2+1}}(C_{\text{%
F,AF,cl}}^{\prime }-1+3\ln 3)  \label{FF}
\end{eqnarray}
($x<1$)$.$ The constants $C_{\text{F,AF,cl}}$ and $C_{\text{F,cl}}^{\prime }$
can be in principle obtained from numerical calculations or by comparing
with experimental data (see below). However, one should expect that they are
small enough and can be neglected.

\section{Comparison with experimental data}

To discuss the experimental situation, we consider first the compounds with
layered perovskite structure. The parameters used are given by Table 1. The
experimental values of transition temperatures are also given and compared
with the theoretical one (for experimental data see Ref. \cite{Joungh} and
references therein, and Ref.\cite{Birg1} for La$_2$CuO$_4$).

Table 1. The experimental parameters and ordering temperatures of layered
magnets and the corresponding calculated values of $T_M$ in the standard
spin-wave-theory (SWT), SSWT and RPA (in brackets - with account of the
constant $C_{\text{AF}}=-0.7$).

\begin{tabular}{||l|c|c|c|ccccc||}
\hline\hline
Compound & $S$ & $J,$K & $J^{\prime },$K & \multicolumn{1}{|c|}{$\Delta _0$}
& \multicolumn{1}{c|}{$T_M^{\text{SWT}},$K} & \multicolumn{1}{c|}{$T_M^{%
\text{SSWT}},$K} & \multicolumn{1}{c|}{$T_M^{\text{RPA}},$K} & $T_M^{\text{%
exp}}$, K \\ \hline\hline
La$_2$CuO$_4$ & 1/2 & 1600 & 0.8 & \multicolumn{1}{|c|}{$\approx $0} & 
\multicolumn{1}{c|}{672} & \multicolumn{1}{c|}{537} & \multicolumn{1}{c|}{343
} & 325 \\ 
K$_2$NiF$_4$ & 1 & 102 & $\approx $0 & \multicolumn{1}{|c|}{0.0088} & 
\multicolumn{1}{c|}{160} & \multicolumn{1}{c|}{125} & \multicolumn{1}{c|}{
90.0\ (97.0)} & 97.1 \\ 
Rb$_2$NiF$_4$ & 1 & 82 & $\approx $0 & \multicolumn{1}{|c|}{0.046} & 
\multicolumn{1}{c|}{180} & \multicolumn{1}{c|}{118} & \multicolumn{1}{c|}{
88.4 (95.0)} & 94.5 \\ 
K$_2$MnF$_4$ & 5/2 & 8.4 & $\approx $0 & \multicolumn{1}{|c|}{0.015} & 
\multicolumn{1}{c|}{74.8} & \multicolumn{1}{c|}{52.1} & \multicolumn{1}{c|}{
42.7 (45.1)} & 42.1 \\ 
CrBr$_3$ & \multicolumn{1}{c|}{3/2} & \multicolumn{1}{c|}{12.38} & 
\multicolumn{1}{c|}{1.0} & \multicolumn{1}{c|}{0.024} & \multicolumn{1}{c|}{
79.2} & \multicolumn{1}{c|}{51.2} & \multicolumn{1}{c|}{39.0} & 40.0 \\ 
\hline\hline
\end{tabular}

\noindent The values of $J^{\prime }$ and $\Delta _0=\Delta (0)$ are
obtained from the low-temperature behavior of the sublattice magnetization.
This procedure gives a possibility to determine correctly the parameters
since the results obtained in Sects. \ref{TempDep} and \ref{E-ax} work well
at low temperatures (in particular they give correct results for
ground-state renormalizations). Note that for the anisotropic perovskites
the parameters obtained are also in agreement with the experimental data on
the spin-wave spectrum \cite{Joungh}. It should be stressed that the
experimentally observable gap in the spin-wave spectrum is $\Delta (T)$
whereas $D,\eta $ plays the role of the bare parameters. The same situation
takes place for the in the quasi-2D case where $J\gamma /S$ and $J^{\prime
}\gamma ^{\prime }/S$ are experimentally observable rather than the bare
parameters $J$ and $J^{\prime }.$ Since in the systems under consideration
the parameters $\gamma ^{\prime }$ and $\Delta $ have strong temperature
dependence (see Sects. \ref{TempDep} and \ref{E-ax}) it is important to take
into account this dependence when treating the experimental data.

One can see from the Table 1 that for all the systems the estimated values
of transition temperatures are close to the experimental results (for La$_2$%
CuO$_4$ the experimental data on $J^{\prime }$ are contradictionary; one of
the possibilities to improve the agreement with experimental data is
introducing small easy-axis anisotropy \cite{OurRG}). At the same time,
using SSWT without fluctuation corrections overestimates $T_N$ by about 1.7
times, although improves somewhat the results of standard spin-wave theory.
For the anisotropic compounds $T_N$ is slightly underestimated. This may be
due to two reasons: non-zero value of $C_{\text{AF}}$ in this case ($C_{%
\text{AF}}\simeq -0.7$ is obtained by best fit to the experimental data) and
(less important) small interlayer coupling which also increases the
transition temperature.

The temperature dependence of the sublattice magnetization for K$_2$NiF$_4$
is shown and compared with different theoretical results in Fig. \ref
{FigExprt}. The regimes (i),(ii) correspond to $T<80$K where the RPA (RG)
result is in good agreement with the experimental data. It can be seen also
that the account of the function $\Phi _a^{\text{AF}}(t/\overline{\sigma })$
given by (\ref{FF}) improves considerably the agreement in the crossover
temperature region. The $1/N$-expansion in the $O(N)$ model\cite{Our1/N}
also gives satisfactory description of this region, but does not describe
correctly low enough temperatures, since, as discussed in the Introduction,
it implies an essentially non-spin-wave picture of the excitation spectrum.

The parameter values and Curie temperature for the ferromagnetic compound
CrBr$_3$ (see Ref. \cite{CrBr3} for experimental data) are also presented in
Table 1. Here both interplane coupling and anisotropy are important, and we
obtain (see also Ref. \cite{OurRG}) 
\begin{eqnarray}
T_M &=&4\pi |J|\overline{S}_0\gamma _0  \nonumber \\
&&\ \ \ \ \ \times \left\{ \ln [2q_0^2/(\Delta _c+2\alpha _c+\sqrt{\Delta
_c^2+4\alpha _c\Delta _c})]+2\ln \frac{4\pi |J|\overline{S}_0\gamma _0}{T_M}%
+C\left( \frac \Delta \alpha \right) \right\} ^{-1}  \label{tTc}
\end{eqnarray}
where 
\begin{eqnarray}
\alpha _c &=&J^{\prime }\gamma _c^{\prime }/(J\gamma _c)=T_M|J^{\prime
}|/\left[ 4\pi (J\gamma _0)^2\right]  \nonumber \\
\Delta _c &=&\Delta _0\left[ T_M/(4\pi J\overline{S}_0\gamma _0)\right] ^2
\end{eqnarray}
The (unknown) function $C(\Delta /\alpha )$ satisfies $C(0)=C^{\prime }$ and 
$C(\infty )=C$ where $C$ and $C^{\prime }$ are defined above. Since the
in-plane lattice structure is non-square, we have used the effective value
of in-plane exchange integral determined in the continuum limit from the
excitation spectrum as $E_{{\bf q}}\simeq \gamma J_{\text{ef}}(q_x^2+q_y^2)+%
{\cal O}(J^{\prime },J\Delta ).$ One can see that the agreement with
experimental $T_C$ is excellent.

\section{Conclusions}

\label{Concl}In the present paper we have investigated in detail the
capabilities of the self-consistent spin-wave theory, which is based on
boson representations for spin operators, for description of layered
magnets. To improve the SSWT at not too low temperatures, we have introduced
a pseudofermion field. For magnets with low transition point, $T_M\ll
|J|S^2, $ analytical results were obtained. These results have different
forms in quantum $(T\ll |J|S)$ and classical $(T\gg |J|S)$ regimes. In the
quantum case the magnetization demonstrates in some temperature region the $%
T\ln T$-behavior. The proposed version of SSWT gives a qualitative (and at
not too high temperatures - quantitative) description of thermodynamics of
layered magnets. An important advantage of SSWT (in comparison with the
methods that are based on investigation of continuum models, e.g.,
non-linear sigma model) is a possibility to describe the short-range order
above the transition point.

At the same time, even in the case of layered magnets SSWT is unable to
treat quantitatively the transition points and thermodynamics at high enough
temperatures ($T>0.8T_M$). We have performed a systematic treatment of
corrections to SSWT in the case $T_M\ll |J|S^2$, which is based on summation
of higher order $1/S$-terms. The inclusion of RPA corrections (which permits
to take into account next-leading logarithmic singularities) yields an
excellent description of the behavior of (sublattice) magnetization at
arbitrary $T<T_M$ except for a narrow critical region, where an account of
non-spin-wave excitations is required. The approach used is somewhat
reminiscent of the theory of itinerant magnets \cite{Moria}. As well as in
the latter case, the fluctuation corrections within the RPA approximation
lead to significant lowering of the transition temperature and improve
radically the agreement with experimental data. The simple analytical
results obtained give a quantitative description of the magnetization
behavior practically up to $T_M$. At the same time, the consideration of the
critical region (e.g., correct calculation of critical exponents) requires
an account of essentially non-spin-wave excitations. For quasi-2D isotropic
magnets this can be performed within the $1/N$-expansion in $O(N)$ model\cite
{Our1/N}. A description of the critical region for magnets with the
easy-axis anisotropy, where the topological (domain-wall) excitations are
present, is still an open problem.

A regular calculation of higher-order corrections in the 3D case, where the
kinematical spin-wave interaction should be also taken into account and the
logarithmic terms are absent, is also the matter for further investigations.
The same is valid for magnets with essential role of topological excitations
(easy-plane systems, antiferromagnetic half-integer spin chains etc.).

\section*{Appendix A. Variational principle in the Heisenberg model}

In this Appendix we consider a variational approach to SSWT. This is a
generalization of approaches of Refs. \cite{BlochVar,Rast2} to quasi-2D
case, which gives a possibility to improve the behavior of the magnetization
at not too small $J^{\prime }/J$.

We apply the Feynman-Peierls-Bogoliubov variation principle for the free
energy $F=-T\ln $Sp$(e^{-\beta H})$ (see, e.g., \cite{Tyab,Feinm}) 
\begin{equation}
F<F_0+\langle H-H_0\rangle _0
\end{equation}
where $H_0$ is the trial Hamiltonian, $F_0$ is the free energy corresponding
to $H_0$, $\langle ...\rangle _0$ stands for the average with $H_0$.

Consider first the case of a ferromagnet. Assuming the spin-wave character
of spin dynamics we choose $H_0$ as the Hamiltonian of non-interacting
bosons and fermions 
\begin{eqnarray}
H_0 &=&-\frac 12J_0\gamma S+\gamma \sum\limits_{{\bf k}}(J_0-J_{{\bf k}})b_{%
{\bf k}}^{\dagger }b_{{\bf k}}^{}+(2S+1)\gamma \,J_0\sum\limits_{{\bf k}}c_{%
{\bf k}}^{\dagger }c_{{\bf k}}^{}  \nonumber \\
&&\ \ \ \ \ -\,\mu \,\sum\limits_{{\bf k}}\left[ b_{{\bf k}}^{\dagger }b_{%
{\bf k}}+(2S+1)c_{{\bf k}}^{\dagger }c_{{\bf k}}\right]
\end{eqnarray}
This expression differs from the Hamiltonian of the standard spin-wave
theory by the factor $\gamma $ which is a variational parameter describing
the renormalization of the spin-wave spectrum, and by the presence of the
Fermi operators taking into account the kinematical interaction between spin
waves. Then we obtain 
\begin{equation}
F_0=T\sum\limits_{{\bf q}}\ln \frac{1-\exp (-E_{{\bf q}}/T)}{1-\exp (-E_f/T)}
\end{equation}
and 
\begin{eqnarray}
\langle \,H\,\rangle _0 &=&-\frac 12J_0\overline{S}^2-\overline{S}%
\sum\limits_{{\bf q}}J_{{\bf q}}N_{{\bf q}}-\frac 12\sum\limits_{{\bf pq}}J_{%
{\bf q-p}}N_{{\bf q}}N_{{\bf p}}-\mu \sum\limits_{{\bf q}}\left[ N_{{\bf q}%
}+(2S+1)N_f\right] ,  \nonumber \\
{\langle }\,H_0\,\rangle _0 &=&\gamma J_0(S-\overline{S})-\gamma
\sum\limits_{{\bf q}}J_{{\bf q}}N_{{\bf q}}-\mu \sum\limits_{{\bf q}}\left[
N_{{\bf q}}+(2S+1)N_f\right] .
\end{eqnarray}
where the magnetization reads 
\begin{equation}
\overline{S}=(S+\frac 12)\coth \frac{E_f}T-\frac 12-\sum\limits_{{\bf q}}N_{%
{\bf q}}
\end{equation}
and $E_f=\gamma J_0-\mu .$ The equation for $\gamma $ is determined from the
condition $\partial F/\partial \gamma =0$ and has the form 
\begin{equation}
(\gamma -\overline{S})\,\left[ \sum\limits_{{\bf q}}J_{{\bf q}}\frac{%
\partial N_{{\bf q}}}{\partial \gamma \,}+J_0\frac{\partial \overline{S}}{%
\partial \gamma \,}\right] =\sum\limits_{{\bf pq}}J_{{\bf p-q}}N_{{\bf p}}%
\frac{\partial N_{{\bf q}}}{\partial \gamma \,}+\sum\limits_{{\bf q}}J_{{\bf %
q}}N_{{\bf q}}\frac{\partial \overline{S}}{\partial \gamma \,}
\label{GammaVF}
\end{equation}

In the antiferromagnetic case we use the trial Hamiltonian 
\begin{eqnarray}
H_0 &=&-\frac 12J_0\gamma S+\frac 12\gamma \sum\limits_{{\bf k}}\left[
J_0(a_{{\bf k}}^{\dagger }a_{{\bf k}}+b_{{\bf k}}^{\dagger }b_{{\bf k}})-J_{%
{\bf k}}(a_{{\bf k}}^{\dagger }b_{-{\bf k}}^{\dagger }+a_{{\bf k}}b_{-{\bf k}%
})\right]  \nonumber \\
&&\ \ +(2S+1)\gamma \,J_0\sum\limits_{{\bf k}}(c_{{\bf k}}^{\dagger }c_{{\bf %
k}}^{}+d_{{\bf k}}^{\dagger }d_{{\bf k}}^{})-\,\mu \,\sum\limits_{{\bf k}%
}\left[ a_{{\bf k}}^{\dagger }a_{{\bf k}}+b_{{\bf k}}^{\dagger }b_{{\bf k}%
}+(2S+1)(c_{{\bf k}}^{\dagger }c_{{\bf k}}+d_{{\bf k}}^{\dagger }d_{{\bf k}%
})\right]
\end{eqnarray}
to obtain 
\begin{equation}
(\gamma -\overline{S})\,\left[ \sum\limits_{{\bf q}}J_{{\bf q}}\frac{%
\partial L_{{\bf q}}}{\partial \gamma \,}+J_0\frac{\partial \overline{S}}{%
\partial \gamma \,}\right] =\sum\limits_{{\bf pq}}J_{{\bf p-q}}L_{{\bf p}}%
\frac{\partial L_{{\bf q}}}{\partial \gamma \,}+\sum\limits_qJ_{{\bf q}}L_{%
{\bf q}}\frac{\partial \overline{S}}{\partial \gamma \,}  \label{GammaVAF}
\end{equation}
where 
\begin{eqnarray}
E_{{\bf q}} &=&\sqrt{(\gamma \,J_0-\mu \,)\,^2-(\gamma \,J_{{\bf q}})\,^2} 
\nonumber \\
L_{{\bf q}} &=&\frac{\gamma J_{{\bf q}}}{E_{{\bf q}}}(N_{{\bf q}}+\frac 12)
\end{eqnarray}
and 
\begin{equation}
\overline{S}=(S+\frac 12)\coth \frac{E_f}T-\sum\limits_{{\bf q}}\frac{%
J_0\gamma -\mu }{E_{{\bf q}}}(N_{{\bf q}}+\frac 12)
\end{equation}
In the 2D case ($J^{\prime }=0$) we have 
\begin{equation}
\sum\limits_{{\bf pq}}J_{{\bf p-q}}N_{{\bf p}}\frac{\partial \,N_{{\bf q}}}{%
\partial \gamma \,}=\frac 1{J_0}\left( \sum\limits_{{\bf q}}J_{{\bf q}}N_{%
{\bf q}}\right) \left( \sum\limits_{{\bf p}}J_{{\bf p}}\frac{\partial \,N_{%
{\bf p}}}{\partial \gamma \,}\right) \   \label{Id}
\end{equation}
and a similar result with the replacement $N_{{\bf q}}$ $\rightarrow $ $L_{%
{\bf q}}$. Then the equations (\ref{GammaVF})\ and (\ref{GammaVAF})\ reduce
to 
\begin{eqnarray}
\gamma -\overline{S} &=&\frac 1{J_0}\sum\limits_{{\bf q}}J_{{\bf q}}N_{{\bf q%
}}\,\,\,\,\,\,\,\,\,\,\,\,\,\,\text{(FM),}  \nonumber \\
\gamma -\overline{S} &=&\frac 1{J_0}\sum\limits_{{\bf q}}J_{{\bf q}}L_{{\bf q%
}}\,\,\,\,\,\,\,\,\,\,\,\,\,\,\,\text{(AFM).}  \label{VI}
\end{eqnarray}
These equations coincide with those of Sect \ref{Q2dBKJ}. At small $%
J^{\prime }/J$ the values of integrals over the Brillouin zone are
determined by the region of small quasimomenta $q_x^2,q_y^2<J^{\prime }/J.$
Then it is possible to put 
\begin{equation}
J_{{\bf p-q}}\simeq J_0,\;J_{{\bf q}}\simeq J_0  \label{a-x}
\end{equation}
in this case, which again leads to Eqs. (\ref{VI}). If $J^{\prime }/J$ is
not small enough, the approximation (\ref{a-x}) becomes invalid and it is
necessary to use the equation (\ref{GammaVF})\ or (\ref{GammaVAF}) up to the
limit of the cubic crystal, $J^{\prime }=J,$ where equality (\ref{Id}) is
again satisfied and passing to (\ref{VI}) becomes justified. The same
situation is realized for hypercubic lattices of any dimensionality in the
nearest-neighbor approximation.

\section*{Appendix B. The solution of integral equation for the renormalized
vertex of spin-wave interaction}

The equation for the vertex (\ref{EqVertex}) has a degenerate kernel. We
search the solution in the form 
\begin{equation}
\Phi ({\bf k,k-q;p-q,p})=J(A{\bf q}-B{\bf p}){\bf k}-2J\widetilde{\Delta }(%
{\bf p,q})
\end{equation}
Then we obtain after some algebraic manipulations 
\begin{eqnarray}
A &=&\frac 2{\overline{S}_0/\gamma -R+M_q}\left[ 1+2{\bf q}({\bf p-q}%
)(\Gamma _q-M_q/q^2)\right.  \nonumber \\
&&\ \ \ \ \ \ \ \left. +2\Delta \left( J\chi _{q0}^{zz}-\Gamma _q\right)
\right]  \nonumber \\
B &=&\frac 2{\overline{S}_0/\gamma -R+M_q}  \nonumber \\
\widetilde{\Delta }({\bf p,q}) &=&\Delta \left[ 1-\Gamma _q(A{\bf pq}+Bq^2%
{\bf )}\right]
\end{eqnarray}
where we have introduced the correction to the magnetization $R=(\overline{S}%
_0-\overline{S})/\gamma $ and the non-uniform susceptibility $\chi
_{q0}^{zz} $ 
\begin{eqnarray}
R &=&\frac T{|J|\gamma ^2}\int_{s<q_0}\frac{d^2{\bf s}}{(2\pi )^2}\frac 1{%
s^2+\Delta }  \nonumber \\
\chi _{q0}^{zz} &=&\frac T{(J\gamma )^2}\int \frac{d^2{\bf s}}{(2\pi )^2}%
\frac 1{(s^2+\Delta )[({\bf s-q})^2+\Delta ]},
\end{eqnarray}
and $\Gamma _q,$ $M_q$ are defined by 
\begin{eqnarray}
\frac T{|J|\gamma ^2}\int \frac{d^2{\bf s}}{(2\pi )^2}\frac{{\bf s}}{%
(s^2+\Delta )[({\bf s-q})^2+\Delta ]} &=&{\bf q}\Gamma _q  \nonumber \\
\frac T{|J|\gamma ^2}\int_{s<q_0}\frac{d^2{\bf s}}{(2\pi )^2}\frac{s_is_j}{%
(s^2+\Delta )[({\bf s-q})^2+\Delta ]} &=&\frac 12R\delta _{ij}-M_q\left( 
\frac{\delta _{ij}}2-\frac{q_iq_j}{q^2}\right)  \label{GM}
\end{eqnarray}
Calculating the integrals in (\ref{GM}) and retaining only terms which are
logarithmically divergent at $\Delta \rightarrow 0\ $yields 
\begin{equation}
\Gamma _q=M_q/q^2=\frac 12|J|\chi _{q0}^{zz}
\end{equation}
where $\chi _{q0}^{zz}$ is given by (\ref{Hiq0}), and finally we get 
\begin{equation}
R=\frac T{2\pi |J|\gamma ^2}\ln \frac{q_0}{\Delta ^{1/2}}
\end{equation}
Combining the above formulas we obtain the result (\ref{FAnsw}) of the main
text.

{\sc CAPTIONS TO FIGURES}

\begin{enumerate}
\item  Calculated temperature dependences of SRO parameters in $S=1/2$
quasi-2D ferro- (right-hand side) and antiferromagnets (left-hand side). The
values of $J^{\prime }/J$ stand at the curves. Dots correspond to 2D case ($%
J^{\prime }/J=0$) without including the pseudofermion contribution,
long-dashed lines and circles to calculations from Eqs. (\ref{gg1}), (\ref
{MagnF}), (\ref{pureAF}) and (\ref{EqI}), (\ref{EqIAF}) respectively for $%
J^{\prime }/J=0.3$. Triangles mark the ordering temperatures\label{FigSROq2D}%
.

\item  Temperature dependence of the (staggered) magnetization $\overline{S}$
for $S=1/2$ quasi-2D ferro- (right-hand side) and antiferromagnets
(left-hand side) with different values of $J^{\prime }/J.\;$Short-dashed
lines show the results without inclusion of pseudofermions, long-dashed
lines in the presence of pseudofermions, and solid lines correspond to the
approximation of effective SRO\ parameter of Sect. \ref{Interp}. For $%
J^{\prime }/J=1$ the solid and long-dashed lines coincide exactly, and for $%
J^{\prime }/J=0.01$ long- and short-dashed lines coincide practically.\label
{FigMagn}

\item  Temperature dependence of the SRO parameters of quasi-2D ferromagnets
for different spin values in the approximation of effective SRO\ parameter%
\label{FigSROs}.

\item  Temperature dependence of the chemical potential of the
boson-pseudofermion systems in quasi-2D ferromagnets in the approximation of
effective SRO\ parameter\label{FigMu}.

\item  Temperature dependence of the magnetization $\overline{S}$ for a $%
S=1/2$ quasi-2D ferromagnet with $J^{\prime }/J=0.1$ in the external
magnetic field\label{FigField}.

\item  Diagrams corresponding to the spin-wave interaction contribution to
different quantities (a) one-particle Green function of SSWT (b) effective
vertex in RPA (c) non-uniform RPA susceptibility (d) correction to
(staggered) magnetization $\delta \overline{S}=S-\overline{S}$\label
{FigFluct}. Simple and bold lines denote the bare and renormalized
one-particle Green functions, point stands for the bare vertex.

\item  Temperature dependence of the relative sublattice magnetization $%
\overline{\sigma }(T)$ of K$_2$NiF$_4$ in SWT, SSWT, RPA approaches and $1/N$
expansion in $O(N)$ model as compared with the experimental data (circles).
The RPA$^{\prime }$ curve corresponds to inclusion of the function $\Phi _a^{%
\text{AF}}(t/\overline{\sigma })$ given by (\ref{FF})\label{FigExprt}.
Short-dashed line is the extrapolation of the result of $1/N$-expansion to
critical region (see Ref. \cite{OurAnis}).
\end{enumerate}


\begin{references}
\bibitem{allen}  A. Allenspach, J. Magn. Magn. Mater. {\bf 129}, 160 (1994).

\bibitem{Birg1}  R.J. Birgeneau et al. Phys. Rev. B{\bf 38}, 6614 (1988)

\bibitem{Birg2}  R.J. Birgeneau, H.J. Guggenheim and G. Shirane, Phys. Rev. B%
{\bf 1}, 2211 (1970).

\bibitem{Birg3}  R.J. Birgeneau et al. Phys. Rev. B{\bf 8}, 304 (1973)

\bibitem{Loly}  P.D. Loly, J. Phys. C{\bf 1}, 1365 (1971)

\bibitem{BlochVar}  M. Bloch, Phys.Rev.Lett, {\bf 9}, 286 (1962).

\bibitem{Rast2}  E. Rastelli, A. Tassi and L. Reatto J.Phys.C{\bf 7}, 1735
(1974).

\bibitem{Arovas}  D.P. Arovas and A.Auerbach, Phys. Rev. B{\bf 38}, 316
(1988).

\bibitem{Yosh}  D.J. Yoshioka, Phys. Soc. Jpn. {\bf 58}, 3733 (1989).

\bibitem{Tak}  M. Takahashi, Phys. Rev. B{\bf 40}, 2494 (1989).

\bibitem{Chakraverty}  S. Chakravarty, B.I. Halperin, and D.R. Nelson,
Phys.Rev.B{\bf 39} 2344 (1989).

\bibitem{Sarker}  S. Sarker, Phys.Rev.B{\bf 40}, 5028 (1989).

\bibitem{Our1st}  V.Yu. Irkhin, A.A. Katanin and M.I. Katsnelson, Phys.
Lett. A{\bf 157}, 295 (1991).

\bibitem{Liu}  Bang-Gui Liu, J. Phys. C{\bf 43}, 8339 (1992).

\bibitem{Barab}  A.F. Barabanov and O.A. Starykh, Pis'ma Zh. Eksp Teor.
Fiz., {\bf 51}, 271.

\bibitem{OurFr}  V.Yu. Irkhin, A.A.Katanin and M.I. Katsnelson, J. Phys. C%
{\bf 4}, 5227 (1992).

\bibitem{Xu}  J.H. Xu and C.S. Ting, Phys. Rev. B, {\bf 42}, 6861 (1990).

\bibitem{Oguch}  T. Oguchi and H. Kitatani J. Phys. Soc. Japan, {\bf 59},
3322 (1990).

\bibitem{Nish}  H. Nishimori and Y. Saika, J. Phys. Soc. Japan, {\bf 59},
4454 (1990).

\bibitem{IK}  V.Yu. Irkhin and M.I. Katsnelson, J.Phys.:Cond.Mat.{\bf 3,}
6439 (1991); see also the paper by V.Yu. Irkhin and M.I. Katsnelson, Z.Phys.
B {\bf 62}, 201 (1986), where a description of LRO in terms of the
singularity of the spin pair correlation function without introducing
anomalous averages was proposed.

\bibitem{Chem}  L.A. Popovich and M.V. Medveded, cond-mat/9803205.

\bibitem{BKJr}  V.G. Baryakhtar, V.N. Krivoruchko, and D.A. Yablonsky,
Zh.Eksp.Teor.Fiz. {\bf 85}, 602 (1983)

\bibitem{BKJBook}  V.G. Baryakhtar, V.N. Krivoruchko, and D.A. Yablonsky, 
{\it Green's Functions in the Theory of Magnetism }[in Russian], Kiev,
Naukova Dumka, 1984.

\bibitem{Our1/N}  V.Yu. Irkhin and A.A. Katanin, Phys.Rev. B{\bf 55}, 12318
(1997).

\bibitem{OurRG}  V.Yu. Irkhin and A.A. Katanin, Phys.Rev. B{\bf 57}, 379
(1998).

\bibitem{OurAnis}  V.Yu. Irkhin and A.A. Katanin, Phys. Lett. A{\bf 232, }%
143 (1997).

\bibitem{Rastelly1}  E. Rastelli and A. Tassi, Phys.Lett. A{\bf 48}, 119
(1974).

\bibitem{Dyson}  F. Dyson, Phys.Rev.{\bf 102, }1217, 1230 (1956).

\bibitem{Oguchi}  T. Oguchi and A. Honma, J.Phys.Soc.Jpn. {\bf 16},{\bf \ }%
79 (1961).

\bibitem{Nagata}  K. Nagata and Y. Tomoto, J.Phys.Soc.Jpn. {\bf 36}, 78
(1974).

\bibitem{Wang}  R.W. Wang and D.L. Millis, Phys.Rev.B{\bf 48}, 3792 (1993).
(1997).

\bibitem{ArovasBook}  A. Auerbach, {\it Interacting\thinspace \thinspace
Electrons\thinspace \thinspace and\thinspace \thinspace Quantum\thinspace
\thinspace Magnetism}, Springer-Verlag, New-York, 1994.

\bibitem{Starykh}  O.A. Starykh, Phys.Rev B {\bf 50}, 16428 (1994).

\bibitem{Starykh1}  O. A. Starykh and A.\ V. Chubukov, Phys.\ Rev. B{\bf 52}%
, 440 (1995).

\bibitem{OurCP}  V.Yu. Irkhin, A. A. Katanin, and M. I .Katsnelson,
Phys.Rev.B{\bf 54, }11953 (1996).

\bibitem{Levanjuk}  A. Levanjuk and N. Garcia, J.Phys.:Cond.Matt. {\bf 4},
10277 (1992); P.A. Serena, N. Garcia, and A. Levanjuk, Phys.Rev.B{\bf 47},
5027 (1993)

\bibitem{Kalash}  V.P. Kalashnikov and M.I. Auslender, Physica A{\bf 100},
443 (1980)

\bibitem{Tyab}  S. V. Tyablikov, {\it Methods in the Quantum Theory of
Magnetism}, Plenum Press, New York, 1967.

\bibitem{LiuSH}  S.H. Liu and D.B. Siano, Phys.Rev. {\bf 164}, 697 (1967).

\bibitem{ChubF}  Yu.A. Kosevich and A.V. Chubukov, Zh.Eksp.Teor.Fiz. {\bf 91}%
, 1105 (1986)

\bibitem{Antz}  T.N. Anzygina and V.S. Slusarev, Fiz. Nizkikh Temperatur, 
{\bf 18}, 261 (1992).

\bibitem{Baxter}  R. J. Baxter, {\it Exactly Solved Models in Statistical
Mechanics,} Academic Press, New York,1982.

\bibitem{Nag}  E.L. Nagaev, {\it Magnetics with Complicated Exchange
Interactions [}in Russian], Nauka, Moscow, 1988.

\bibitem{Brezin}  E. Brezin and J. Zinn-Justin, Phys.Rev.B{\bf 14}, 3110
(1976).

\bibitem{Moria}  T. Moriya, {\it Spin Fluctuations in Itinerant Electron
Magnetism,} Springer, Berlin, 1985.

\bibitem{Damp}  S. Tyc and B.I. Halperin, Phys.Rev.B{\bf 42}, 2096 (1990);
P.Kopeitz and G.Castilla, Phys.Rev.B{\bf 43}, 11100 (1991)

\bibitem{Ma}  S.-K. Ma, {\it Modern Theory of Critical Phenomena, }%
Benjamin-Cummings, Reading, London, 1976.

\bibitem{Vons}  S.V. Vonsovsky, {\it Magnetism}, New York, Wiley, 1974.

\bibitem{Joungh}  {\it Magnetic Properties of Layered Transition Metal
Compounds}, ed. L.J. de Jongh, Cluwer, Dordrecht, 1989.

\bibitem{CrBr3}  H.L. Davis and A. Narath, Phys.Rev.{\bf 134}, A433 (1964).

\bibitem{Feinm}  R.P. Feynman, {\it Statistical Mechanics}, Benjamin, 1972.
\end{references}
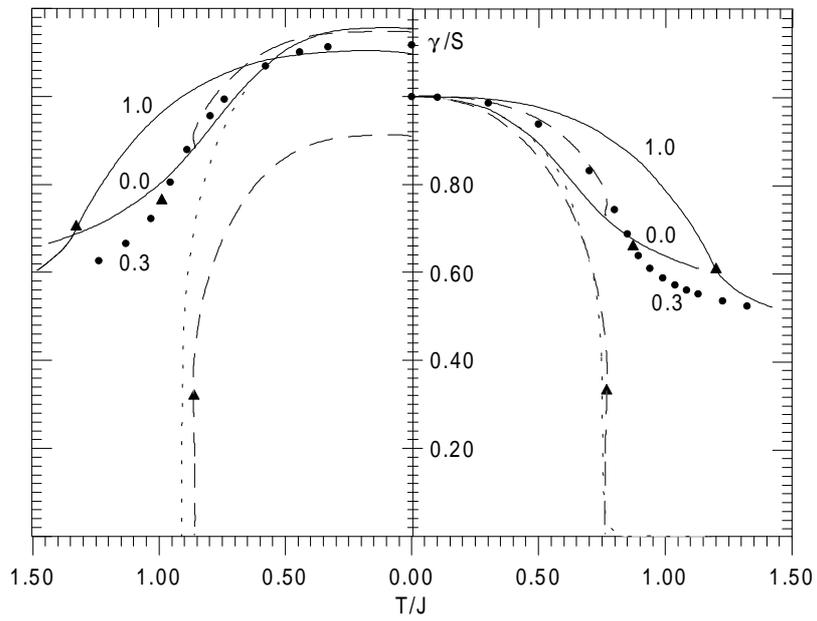
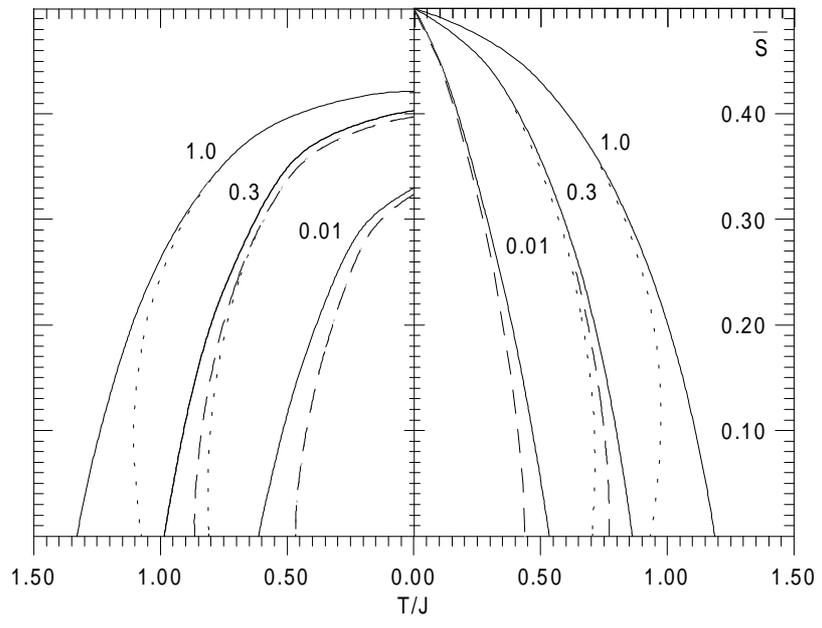
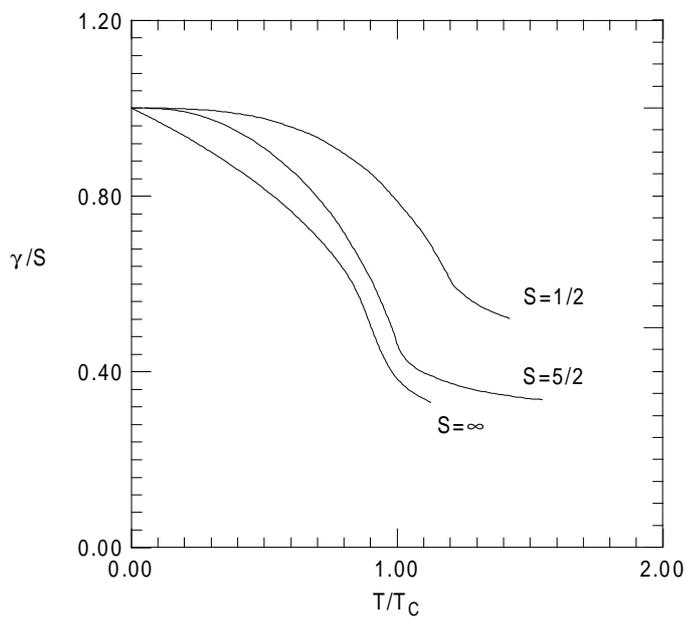
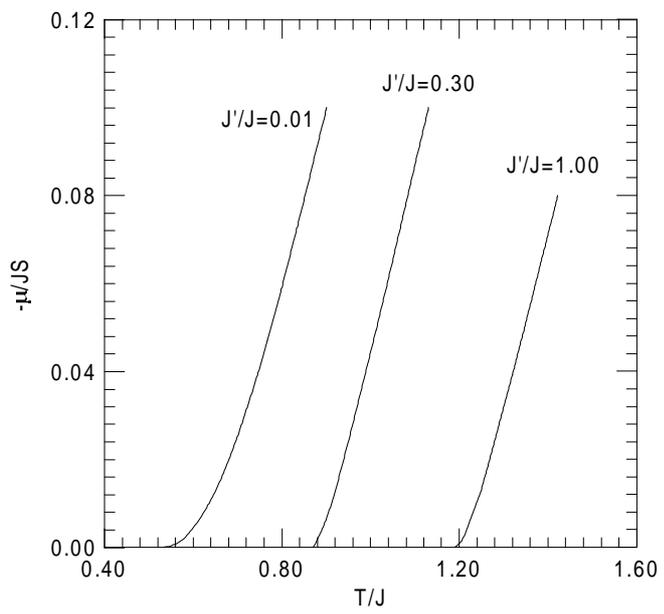
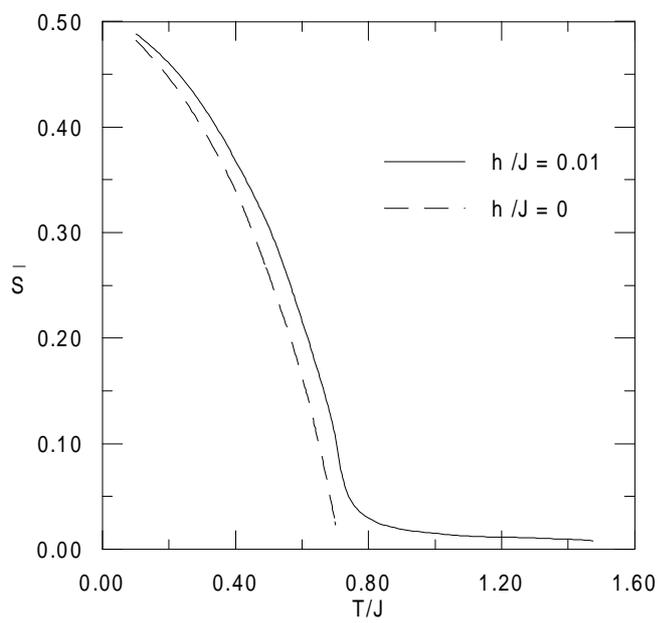
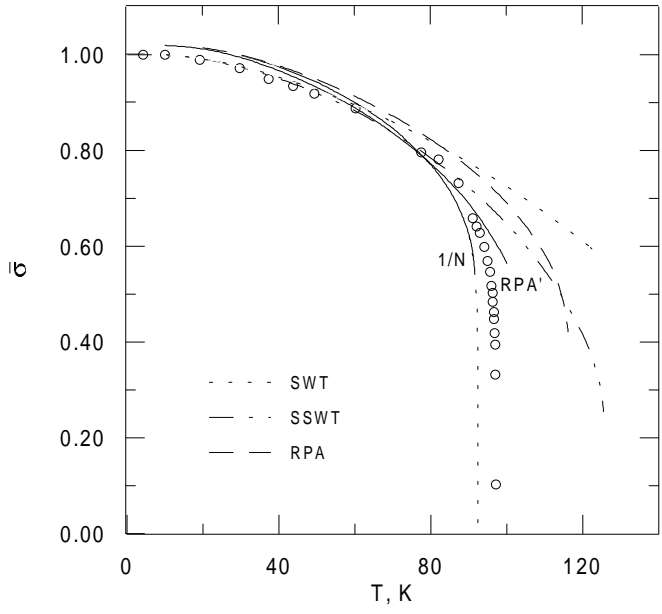
\end{document}